\newcommand{\beq}{\begin{equation}}
\newcommand{\eeq}{\end{equation}}

\newcommand{\bear}{\begin{eqnarray}}
\newcommand{\eear}{\end{eqnarray}}

\newcommand{\tn}{\textnormal}

\documentclass[review]{elsart5p}

\usepackage{graphicx}
\usepackage{enumerate}
\usepackage{pifont}
\usepackage{amssymb}
\usepackage{bbding}
\usepackage{lineno}

\begin{document}

\begin{frontmatter}

\title{Accurate $\gamma$ and MeV-electron track reconstruction with an ultra-low diffusion Xenon/TMA TPC at $10$ atmospheres}
\bigskip
\begin{center} {The NEXT collaboration} \end{center}
\author[1,2,3]{Diego Gonz\'alez-D\'iaz\corauthref{aut1}},
\corauth[aut1]{Corresponding author.}\ead{diegogon@cern.ch}
\author[4] {V. \'Alvarez},
\author[5] {F.I.G. Borges},
\author[10] {M. Camargo},
\author[4] {S. C\'arcel},
\author[1,2] {S. Cebri\'an},
\author[4] {A. Cervera},
\author[5] {C.A.N. Conde},
\author[1,2] {T. Dafni},
\author[4] {J. D\'iaz},
\author[7] {R. Esteve},
\author[5] {L.M.P. Fernandes},
\author[4] {P. Ferrario},
\author[9] {A.L. Ferreira},
\author[5] {E.D.C. Freitas},
\author[6] {V.M. Gehman},
\author[6] {A. Goldschmidt},
\author[4] {J.J. G\'omez-Cadenas\corauthref{aut2}},
\corauth[aut2]{Spokesperson.}\ead{gomez@mail.cern.ch}
\author[10] {R.M. Guti\'errez},
\author[11] {J. Hauptman},
\author[12] {J.A. Hernando Morata},
\author[1,2] {D.C. Herrera},
\author[1,2] {I.G. Irastorza},
\author[13] {L. Labarga},
\author[4] {A. Laing},
\author[4] {I. Liubarsky},
\author[4] {N. Lopez-March},
\author[4] {D. Lorca},
\author[10] {M. Losada},
\author[1,2] {G. Luz\'on},
\author[7] {A. Mar\'i},
\author[4] {J. Mart\'in-Albo},
\author[12] {G. Mart\'inez-Lema},
\author[4] {A. Mart\'inez},
\author[6] {T. Miller},
\author[4] {F. Monrabal},
\author[4] {M. Monserrate},
\author[5] {C.M.B. Monteiro},
\author[7] {F.J. Mora},
\author[9] {L.M. Moutinho},
\author[4] {J. Mu\~noz Vidal},
\author[4] {M. Nebot-Guinot},
\author[6] {D. Nygren},
\author[6] {C.A.B. Oliveira},
\author[17] {J. P\'erez},
\author[15] {J.L. P\'erez Aparicio},
\author[6] {M. Querol},
\author[6] {J. Renner},
\author[14] {L. Ripoll},
\author[4] {J. Rodr\'iguez},
\author[5]{F.P. Santos},
\author[5]{J.M.F. dos Santos},
\author[4] {L. Serra},
\author[6] {D. Shuman},
\author[4] {A. Sim\'on},
\author[16] {C. Sofka},
\author[4] {M. Sorel},
\author[7] {J.F. Toledo},
\author[14] {J. Torrent},
\author[20] {Z. Tsamalaidze},
\author[9]{J.F.C.A. Veloso},
\author[1,2] {J.A. Villar},
\author[16] {R. Webb},
\author[16] {J.T. White \textdagger},
\author[4] {N. Yahlali},

\author[] {}C.
\author[9] {Azevedo},
\author[1,2] {F. Aznar},
\author[18] {D.~Calvet},
\author[1,2] {J. Castel},
\author[18] {E. Ferrer-Ribas},
\author[1,2] {J. A.~Garc\'ia},
\author[18] {I. Giomataris},
\author[1,2] {H. G\'omez},
\author[1,2] {F. J. Iguaz},
\author[1] {A. Lagraba},
\author[18] {A.~Le Coguie},
\author[18] {J.P. Mols},
\author[19] {\"{O}. \c{S}ahin},
\author[1,2] {A. Rodr\'iguez},
\author[1,2] {E. Ruiz-Choliz},
\author[1,2] {L. Segui},
\author[1,2] {A. Tom\'as},
\author[19,3] {R. Veenhof}

\address[1]{Laboratorio de F\'isica Nuclear y Astropart\'iculas, Universidad de Zaragoza, Zaragoza, Spain}
\address[2]{Laboratorio Subterr\'aneo de Canfranc, Canfranc, Spain}
\address[3]{CERN, Geneva, Switzerland}
\address[4]{Instituto de F\'isica Corpuscular (IFIC), CSIC $\&$ Universidad de Valencia, Valencia, Spain}
\address[5]{Departamento de F\'isica, Universidade de Coimbra, Coimbra, Portugal}
\address[10]{Centro de Investigaciones en Ciencias B\'asicas y Aplicadas, Universidad Antonio Nari\~no, Bogot\'a, Colombia}
\address[7]{Instituto de Instrumentaci\'on para Imagen Molecular (I3M), Universitat Polit\`ecnica de Val\`encia, Valencia, Spain}
\address[9]{Institute of Nanostructures, Nanomodelling and Nanofabrication (i3N), Universidade de Aveiro, Aveiro, Portugal}
\address[6]{Lawrence Berkeley National Laboratory (LBNL), Berkeley, USA}
\address[11]{Department of Physics and Astronomy, Iowa State University, Iowa, USA}
\address[12]{Instituto Gallego de F\'isica de Altas Energ\'ias (IGFAE), Univ.\ de Santiago de Compostela, Santiago de Compostela, Spain}
\address[13]{Departamento de F\'isica Te\'orica, Universidad Aut\'onoma de Madrid, Madrid, Spain}
\address[17]{Instituto de F\'isica Te\'orica (IFT), UAM/CSIC, Madrid, Spain}
\address[15]{Dpto.\ de Mec\'anica de Medios Continuos y Teor\'ia de Estructuras, Univ.\ Polit\`ecnica de Val\`encia, Valencia, Spain}
\address[14]{Escola Polit\`ecnica Superior, Universitat de Girona, Girona, Spain}
\address[16]{Department of Physics and Astronomy, Texas A\&M University, Texas, USA}
\address[20]{Joint Institute for Nuclear Research (JINR), Dubna, Russia}
\address[19]{Department of Physics, Uluda\u{g} University, Bursa, Turkey}
\address[18]{IRFU, Centre d'\'Etudes Nucl\'eaires (CEA), Saclay, France}

\begin{abstract}
We report the performance of a 10 atm Xenon/trimethylamine time projection chamber (TPC) for the detection of X-rays (30 keV) and $\gamma$-rays (0.511-1.275 MeV) in conjunction with the accurate tracking of the associated electrons. When operated at such a high pressure and in 1\%-admixtures, trimethylamine (TMA) endows Xenon with an extremely low electron diffusion ($1.3\pm 0.13$ mm-$\sigma$ (longitudinal), $0.8\pm 0.15$ mm-$\sigma$ (transverse) along 1~m drift) besides forming a convenient `Penning-Fluorescent' mixture. The TPC, that houses 1.1 kg of gas in its fiducial volume, operated continuously for 100 live-days in charge amplification mode. The readout was performed through the recently introduced microbulk Micromegas technology and the AFTER chip, providing a 3D voxelization of 8mm$\times$8mm$\times$1.2mm for approximately 10 cm/MeV-long electron tracks. Energy resolutions at full width half maximum (FWHM) inside the fiducial volume ranged from $14.6\%$(30keV) to $4.6\%$(1.275MeV).

This work was developed as part of the R\&D program of the NEXT collaboration for future detector upgrades in the search of the neutrino-less double beta decay ($\beta\beta0\nu$) in $^{136}$Xe, specifically those based on novel gas mixtures. Therefore we ultimately focus on the calorimetric and topological properties of the reconstructed MeV-electron tracks. In particular, the obtained energy resolution has been decomposed in its various contributions and improvements towards achieving the $1.4\%\sqrt{1 \tn{MeV}/\varepsilon}$ levels obtained in small sensors are discussed.
\end{abstract}

\end{frontmatter}


\section{Introduction}
\label{intro}

The ability to perform $\gamma$-ray spectroscopy in conjunction with the accurate electron(s) reconstruction is of contemporary interest in fundamental science and technology. Such an asset is of utmost importance for instance in some implementations of Compton cameras either in standard \cite{AzeCC} or electron-tracking mode \cite{Matsu}, $\gamma$ and X-ray polarimetry \cite{HARPO, Black} as well as nuclear physics \cite{NEXT, AGATA}.
Generally speaking, a `dream' scenario for detection is that where full event containment can be achieved in a medium with sparse ionization;
an optimal (`intrinsic') energy resolution emerges from the subsequent sub-Poisson fluctuations of the released charge \cite{Alzakhov}, while the sense of the momentum vector can be obtained through resolving distinct track features such as characteristic X-ray emission at production and Bragg peak \cite{Gotthard}. Reconstruction of the track's direction requires multiple scattering to be minimized and fidelity to the primary ionization trail to be ensured by proper design choices.
If such is the case, accurate information about the photon energy, direction and sense of its momentum vector and polarization can be retrieved from the emerging lepton tracks, involving in general multi-site cluster reconstruction algorithms. For very rare nuclear $\beta$-decays, on the other hand, an extremely high selectivity and background suppression can be achieved from the combined calorimetric and topological information \cite{NEXT}.

\section{High Pressure Xenon TPCs for MeV-$\gamma$'s and electrons, and the case of NEXT}
\label{intro}

High pressure Xenon time projection chambers (TPCs) capable of simultaneously reconstructing the energy and trajectory of MeV-electrons to great accuracy have been  extensively discussed recently, and the reader is referred to \cite{DaveTPC, LOI} for details. As a gas, Xenon allows the initial ionization to span over a considerable distance even at high pressure ($\sim\!\!10$~cm/MeV at 10~bar) expanding the track details and facilitating its reconstruction. At the same time, lepton tracks can be fully contained up to energies of the order of 10 MeV for a 100 kg/10 bar detector, by virtue of the high stopping power of Xenon. The scale of a cylindrical 1 ton/10 bar detector with a height/diameter aspect ratio $h/\phi=1$ approaches the asymptotic mean free path for pair production $h=\lambda_{e+e-}=9/7X_o\sim 2$ m, ensuring a high $\gamma$ detection efficiency for all energies and event containment up to 10's of MeV.\footnote{$X_o$ is the radiation length.}

The excellent tracking prospects of gaseous Xenon come at a modest $\times 2.5$ worsening of the energy resolution relative to that of state of the art Germanium detectors:

\beq
\mathfrak{R}_0 = 2.35\sqrt{\frac{F W_I}{\varepsilon}}=0.45\% \sqrt{\frac{1 \tn{MeV}}{\varepsilon}} \label{resInt}
\eeq
I.e., for a fixed energy deposit ($\varepsilon$), resolution is dictated by the Fano factor $F$ ($0.15\pm0.02$) and the average energy needed to create an electron-ion pair ($W_I=24.8$ eV), \cite{DaveTPC}.
In practical detectors this value can be approached through secondary light amplification (electroluminescence) \cite{NEXT-DBDM}, a process that benefits from pressurization \cite{Policarpo}. 
 For the simultaneous reconstruction of the MeV-electron track, the separate optimization of the tracking (eg. with SiPMs) and calorimetric function (PMTs)
has been pursued and demonstrated by the NEXT collaboration \cite{NEXT_TDR, DEMO, Lorca}.

NEXT detection technique, based on the accurate reconstruction of the 2-electron tracks emerging from the $^{136}$Xe $\beta\beta0\nu$ decay with a total energy equaling $Q_{\beta\beta}=2.457$ MeV, aims at suppressing the bulk of $\gamma$-ray backgrounds from natural radioactivity with rejection factors of $\sim\!10^7$, as required for the exploration of the inverted hierarchy of neutrino masses.
At such a sensitivity level, achieving background-free conditions (if possible at all) requires advances on the topological analysis beyond the double Bragg-peak (double `blob') identification long championed by the Gotthard collaboration \cite{Gotthard}. Arguably, the detailed tracking in an axial magnetic field may provide a new path \cite{JJBaTa}: in this case $\gamma$-backgrounds display 1 blob and 1 definite curvature (or chirality), while a $\beta\beta0\nu$ signal will double these features. For a relativistic track propagating over a distance $l$, the kick angle and the angular straggling projected into the bending plane can be approximated, respectively, by \cite{pdg}:

\bear
&\theta_{B}&=17^\circ\frac{l}{1\tn{mm}}\frac{B}{1\tn{T}} \times \frac{1\tn{MeV}}{\varepsilon} \label{B} \\
&\theta_{M-\!S}&=11^\circ\sqrt{\frac{l}{1\tn{mm}}} \times \sqrt{\frac{P}{10 \tn{bar}}} \times \frac{1\tn{MeV}}{\varepsilon} \label{MS}
\eear
For practical magnetic-fields ($B$) and pressures ($P$) this sets the desirable tracking step at the mm-level, below which multiple-scattering overrules magnetic bending.\footnote{It is later shown that the spatial extent of an MeV-electron blob at 10~bar is 2~mm ($1$-$\sigma$). The suitability of mm-tracking will be demonstrated elsewhere, with the support of detailed simulations.}
Extracting additional topological information in the $\beta\beta0\nu$ quest therefore implies that any blurring introduced by the diffusion of the initial ionization or the instrumental response should be minimized accordingly. Since this is not viable in pure Xenon, with an electron diffusion at the scale of 1~cm for 1~m drift \cite{NEXT-DBDM}, the possibility of a new generation of Xe-TPCs with a 10-fold increased track sharpness is being explored. Achieving this goal, while preserving both the near-intrinsic calorimetric performance and the primary scintillation at the m-scale, presents a formidable challenge.

\section{Trimethylamine and its `magic' mixtures with Xenon}

It has been noted that Xenon, when mixed with Trimethylamine (TMA: N(CH$_3$)$_3$) may, besides reducing the electron diffusion, form a mixture capable of simultaneously displaying Penning effect and scintillation \cite{DavePF}. This concept is dubbed, hereafter, `Penning-Fluorescent'. The idea that Xenon VUV-light may be efficiently re-absorbed in fluorescent and ionization degrees of freedom of the additive has appealing implications, allowing an improvement of the energy resolution through the reduction of the Fano factor and the increase of the ionization response $1/W_I$, as well as wavelength-shifting the primary and secondary scintillation. The possibility of using TMA for fine-tuning recombination in the search for directional dark-matter signals \cite{HerreraReco} as well as for charge neutralization of Ba$^{++}$ to Ba$^{+}$ (central to some Ba-tagging schemes, \cite{JJBaTa, Sinclair}) adds further value to this molecule, however these sophisticated scenarios (although qualitatively verified) remain speculative at the moment and are not discussed here.

\begin{figure}[hb!!!]
 \centering
 \includegraphics*[width=7.5cm]{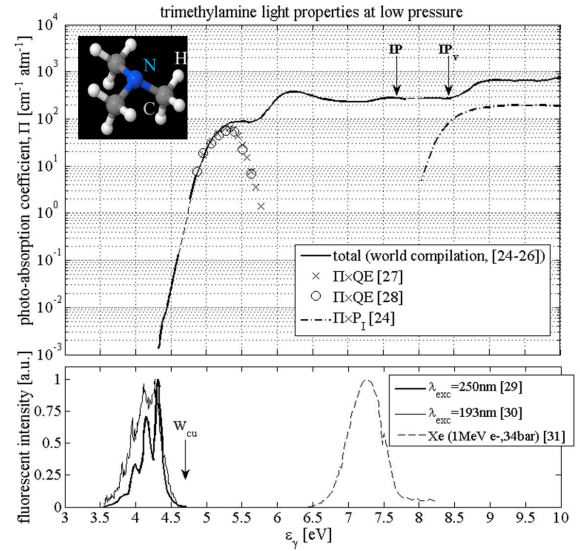}
 \caption{Top panel: World data compilation of TMA light properties at low pressure (1-100 mbar) for the photo-absorption coefficient, after Grosjean\cite{Grosjean}, Tannembaum\cite{Tannembaum} and Halpern\cite{Halpern}. Fluorescent quantum efficiency (QE) included as provided by Obi\cite{Obi1} (crosses) and Cureton\cite{Cureton} (circles), photo-ionization probability from Grosjean\cite{Grosjean} (dot-dashed). Bottom panel: TMA and Xe scintillation spectra for different excitation mechanisms. The spectrum for 250 nm excitation \cite{ObiXe} (QE=100\%) is shown with thick line together with that for 193 nm excitation as obtained by Cardoza in \cite{Cardoza} (thin). Xenon scintillation as measured by Koehler\cite{Koehler} for MeV-electrons at high pressure is indicated by dashed lines. Inset with molecule in 3D from NIST.}
 \label{TMAlight}
 \end{figure}

Trimethylamine is the smallest tertiary amine, and it is known to display a strong (and relatively fast) fluorescence from its lowest Rydberg 3s state ($\tau_{3s}=44$ ns), with quantum yields up to 100\% depending on the excitation wavelength $\lambda_{exc}$. As for the case of ammonia, a large Stokes shift arises due to the transition between a pyramidal and planar geometry during the photon absorption-emission process. Hence, self-transparency can be anticipated for 1\% admixtures with noble gases up to 10's of meters at 10 bar (Fig. \ref{TMAlight}). Contrary to ammonia, the photo-absorption coefficient of TMA does not present marked vibrational bands, a consequence of the smearing introduced by the methyl rotor groups. This fact provides, on the other hand, additional electron cooling and so electron diffusion coefficients as low as $D_{T,L}^*=250$-$350$ $\mu \tn{m}/\sqrt{\tn{cm}}\times\sqrt{\tn{bar}}$ have been reported in \cite{DiegoTMA}. Such a diffusion would imply for a 10 bar detector a spread of the initial ionization at the scale of 1 mm-$\sigma$ over a 1 m path, as confirmed by this work, representing the smallest electron diffusion of any existing TPC that is known to the authors.

The TMA fluorescence spectrum is centered at around 300 nm and conveniently ends at the work function of copper (Fig. \ref{TMAlight}-down) thus reducing feedback from the photo-effect at metal surfaces. TMA itself is strongly opaque to Xenon light, with a mean free path of 0.4 mm for a 1\% TMA admixture at 10 bar.\footnote{Importantly, Xenon emission sits below the vertical ionization potential of TMA (Fig. \ref{TMAlight}), suppressing the track blurring that would be originated by photo-effect in the gas.} With additional VUV-quenching provided during the Xe-excimer formation stage (e.g., through the scintillating, Penning and photo-dissociative pathways made available by TMA) stable charge amplification up to several $1000$'s can be achieved for pressures not exceeding 5~bar \cite{DiegoTMA,XeTMA,Ramsey}. The gain enhancement relative to that in pure Xenon has been often interpreted as a consequence of the close proximity of the TMA ionization potential to the energy of the excited states of Xe species, i.e., to the Penning effect.

In this work it is described, mainly throughout section \ref{setup}, the technical behaviour of a Xe-TMA TPC based on a specific charge-amplifying structure (Micromegas). We later extend the experimental survey on electron swarm properties of Xe-TMA mixtures at 10~bar in section \ref{SwarmSec}, illustrating their impact in the reconstruction of extended tracks in section \ref{ExtTracks}. Section \ref{disc} is devoted to a discussion on the calorimetric response and the Penning-Fluorescent characteristics.

\section{Experimental setup: NEXT-MM}\label{setup}
\subsection{System description and Micromegas readout plane}

NEXT-MM is a 73 liter cylindrical vessel made of stainless steel, equipped with a field cage and $8$ mm$\times 8$ mm pixelated Micromegas anode, assembled and read out in TPC fashion. It has been developed by the NEXT collaboration to assess the performance of this family of micro-patterned charge readouts on large areas (700~cm$^2$) and high pressure, so as to accurately study MeV-electron tracks in low-diffusion mixtures for realistic conditions of event containment. The TPC constituent materials are chosen to be radiopure: copper, PEEK, Teflon\texttrademark~ and polyimide, the main exception being the (solder-less) multi-pin signal connectors based on LCP.\footnote{Liquid crystal polymer. `All-in-one' sensor+cable designs combined with shielding is the most readily available solution to this persistent problem.} The Micromegas were specially manufactured at CERN with the `microbulk' technique \cite{muBulk} provided it grants a radioactivity content below 0.4~Bq/m$^2$ for the $^{238}$U and $^{232}$Th natural chains \cite{HecRP}. Their progeny ($^{214}$Bi and $^{208}$Tl) constitutes the main background in $^{136}$Xe $\beta\beta0\nu$ experiments \cite{NEXT_TDR, Topo}. Besides the presence of structural copper, the TPC anode consists of a modest mass of 3~g/m$^2$ of polyimide, from which much lower activities can be expected upon optimization of the fabrication process.

\begin{figure}[hb!!!]
 \centering
 \includegraphics*[width=7cm]{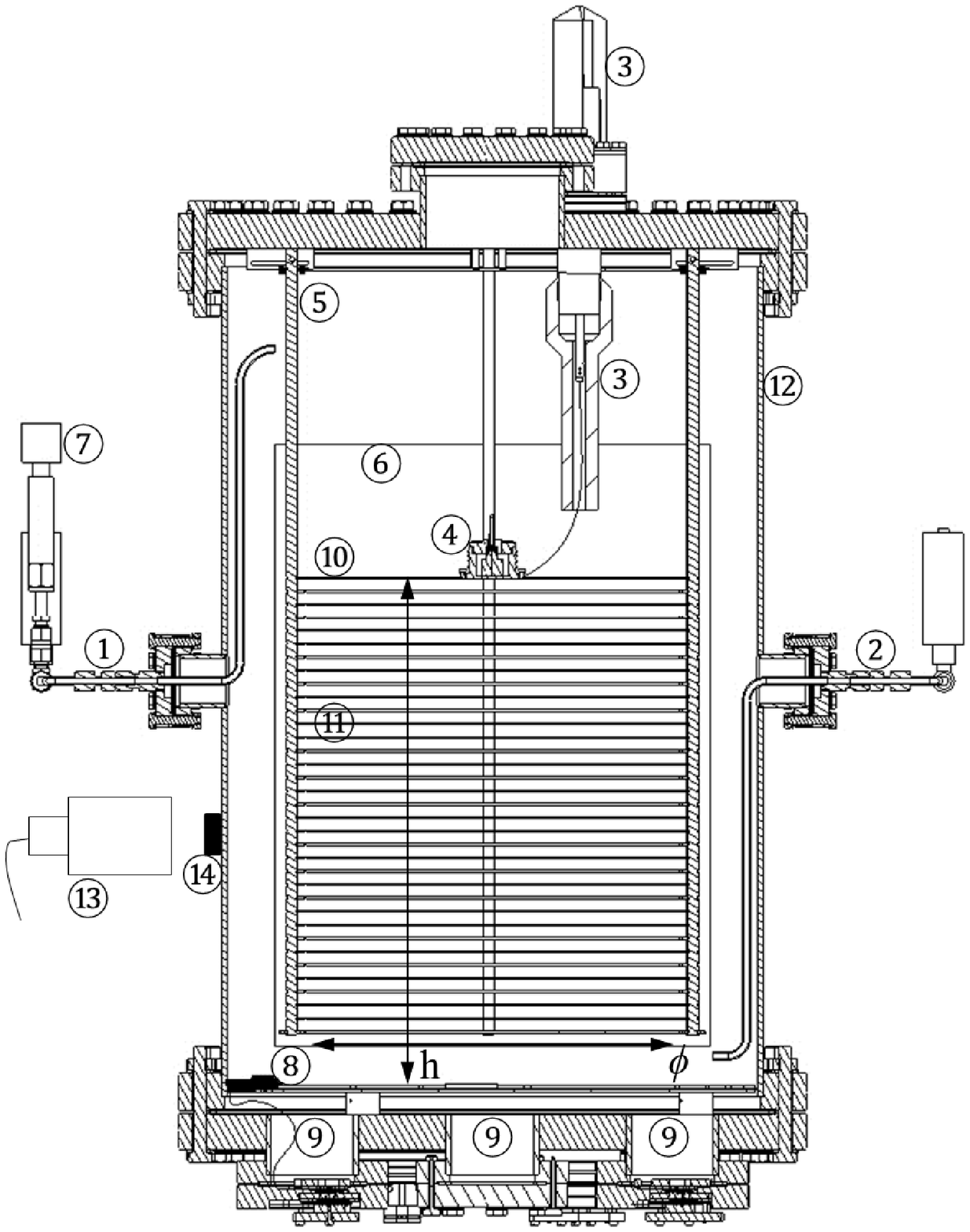}
 \includegraphics*[width=7cm]{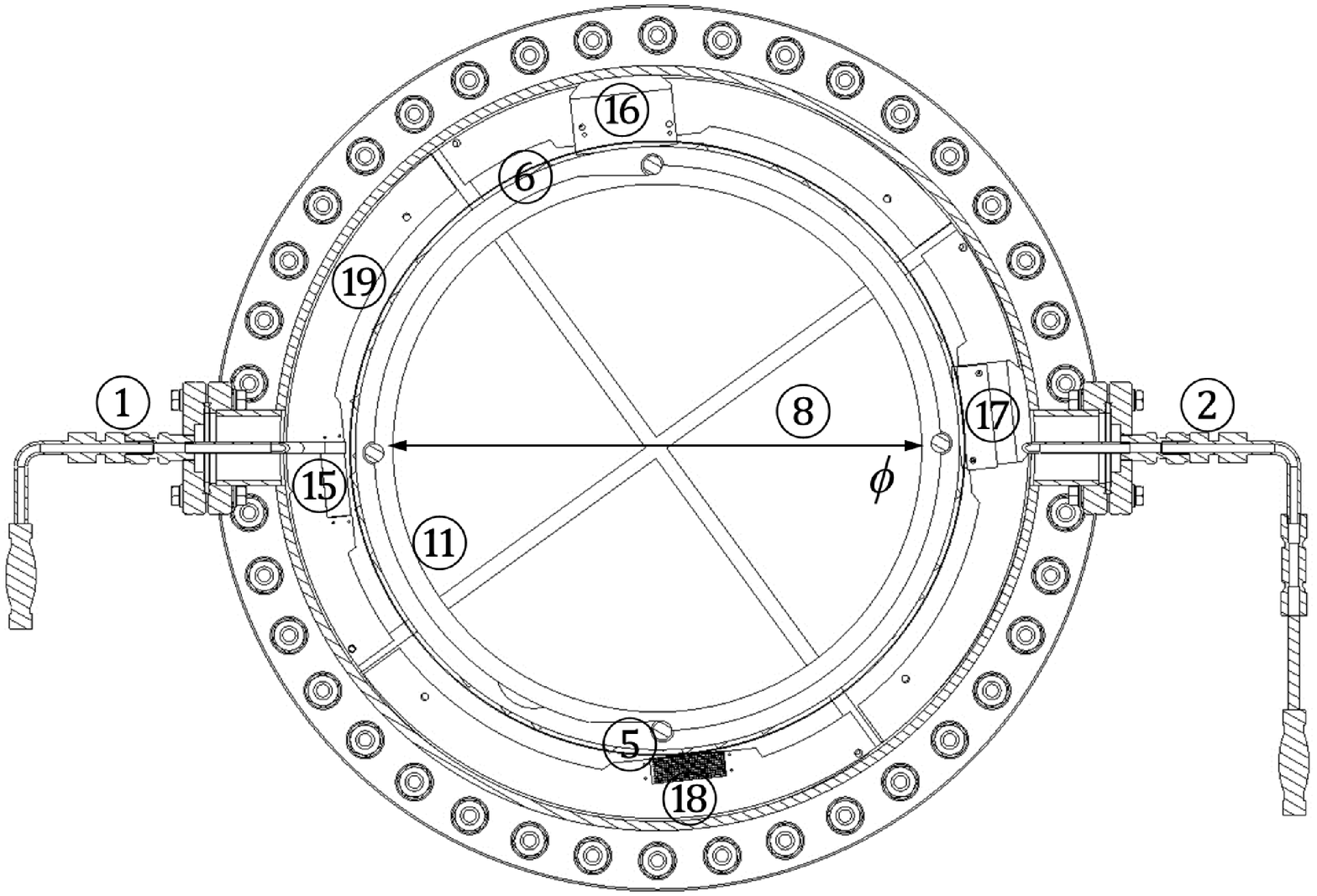}
 \caption{Top panel: cross-section of the NEXT-MM vessel along the direction of gravity (downgoing) with its main elements being labelled. {\large \textcircled{\small 1}} Gas inlet; {\large \textcircled{\small 2}} gas outlet; {\large \textcircled{\small 3}} high voltage (HV) Teflon/copper feed-through; {\large \textcircled{\small 4}} Teflon cage and $^{241}$Am source; {\large \textcircled{\small 5}} PEEK bars for field-cage support; {\large \textcircled{\small 6}} insulating  Cirlex\texttrademark ~rolled sheet; {\large \textcircled{\small 7}} pressure gauge; {\large \textcircled{\small 8}} Micromegas readout plane; {\large \textcircled{\small 9}} flat cable feed-throughs; {\large \textcircled{\small 10}} TPC cathode; {\large \textcircled{\small 11}} field cage rings;  {\large \textcircled{\small 12}} vessel; {\large \textcircled{\small 13}} NaI(Tl) crystal and PMT;  {\large \textcircled{\small 14}} $^{22}$Na source. Bottom panel: up-view of the anode plane from about mid-drift. {\large \textcircled{\small 15}} Footprint for signal out-routing; {\large \textcircled{\small 16}} after flat cable connection; {\large \textcircled{\small 17}} after connection reinforcement; {\large \textcircled{\small 18}} multi-pin SAMTEC connector before assembly; {\large \textcircled{\small 19}} Micromegas mechanical reinforcement. The height of the field cage is $h=38$ cm and the active diameter of the readout plane left by the rings' shadow is $\phi=28$ cm.}
 \label{NEXTMMsections}
 \end{figure}

A cross-section of the vessel together with a close-up of the anode region are shown in Fig. \ref{NEXTMMsections}. The TPC has been designed to house about 1 kg of Xenon in its active volume when operated at 10 bar: $h\times\phi = 38$ cm$ \times 28$ cm. A detailed description of the main TPC components, pressure, vacuum and recovery gas lines can be found in \cite{XeTMA} and \cite{comm}.

Technically, the TPC readout plane was made through a 4-fold sectorial arrangement with each sector enjoying the largest Micromegas manufacturable in the microbulk to date ($\sim\!700/4$~cm$^2$). The construction procedure relies on the chemical etching of doubly copper-clad polyimide foils, being performed in the present case on a 100 $\mu$m square pattern with 40 $\mu$m diameter holes placed at the square's vertexes. Within 50 $\mu$m of the holes' entrance, corresponding to the thickness of the polyimide layer through which the gas amplification takes place, the second copper plate (later acting as anode) is etched with a pattern of $8$mm$\times8$mm induction `pixels'. The number of amplification holes nears 7 million.

For optimization of the Micromegas operation under Xenon-TMA we relied on ground-breaking work performed on small readouts in \cite{XeTMA}. The best energy resolution there obtained was $1.4\% \sqrt{1\tn{MeV}/\varepsilon}$ for 22~keV X-rays (FWHM), with a maximum operating gain of 400 at 10 bar.

\subsection{Data taking and analysis} \label{analysis}

Electron tracks used throughout this work were generated by means of $\gamma$-sources. The released primary ionization drifts towards the TPC anode where it avalanches inside the Micromegas holes, inducing signals on its pixelated anode that can be recorded with specific sampling-ADC electronics, presently based on the AFTER chip \cite{AFTER}. In the absence of light sensors inside the TPC that would be capable of recording the initial scintillation, the $T_o$-information was provided by radiation emitted synchronously from the source, triggering the data acquisition system. For pressures around atmospheric the trigger signal was obtained with the help of a $^{241}$Am source, the $\alpha$-particles from its decays being conveniently tagged by a silicon diode in close proximity. Such a trigger system was encapsulated in a Teflon enclosure above the cathode and served as a source of quasi-simultaneous $^{237}$Np X-rays in the range $25$-$60$ keV, emitted through an opening in the cathode itself (Fig. \ref{NEXTMMsections}-{\large \textcircled{\small 4}}). A $^{22}$Na source located at about mid-drift was used for the measurements at 10~bar (Fig. \ref{NEXTMMsections}-{\large \textcircled{\small 14}}, $\gamma$-ray energies: $0.511$, $1.275$ MeV).

The experimental procedure started with a bake-out cycle of the vessel, gas injection and purification through recirculation, and electric field conditioning. Typically, this lasted about 2 weeks. Following that and prior to the start of long data takings, the operating gain and drift field were systematically scanned for optimal working conditions: $E_{drift}=145$ V/cm/bar, $\tn{gain}\simeq 2000$ (at 1 bar); $E_{drift}=80$ V/cm/bar, $\tn{gain}\simeq 200$ (at 10 bar). Those were found to provide the best balance between Micromegas and field cage stability, highest signal-to-noise (S/N) ratio, convenient drift times, low recombination and negligible attachment. It had been found in earlier works that TMA admixtures around 1\% are convenient both in terms of Micromegas gain response \cite{XeTMA} and reduced diffusion \cite{DiegoTMA} and so we concentrated on such a range. On the other hand, such concentrations would plausibly allow extracting the maximum light yield from the primary TMA scintillation as well as maximizing Penning (section \ref{conc}).

 \begin{figure}[ht!!!]
 \centering
 \includegraphics*[width=7.1cm]{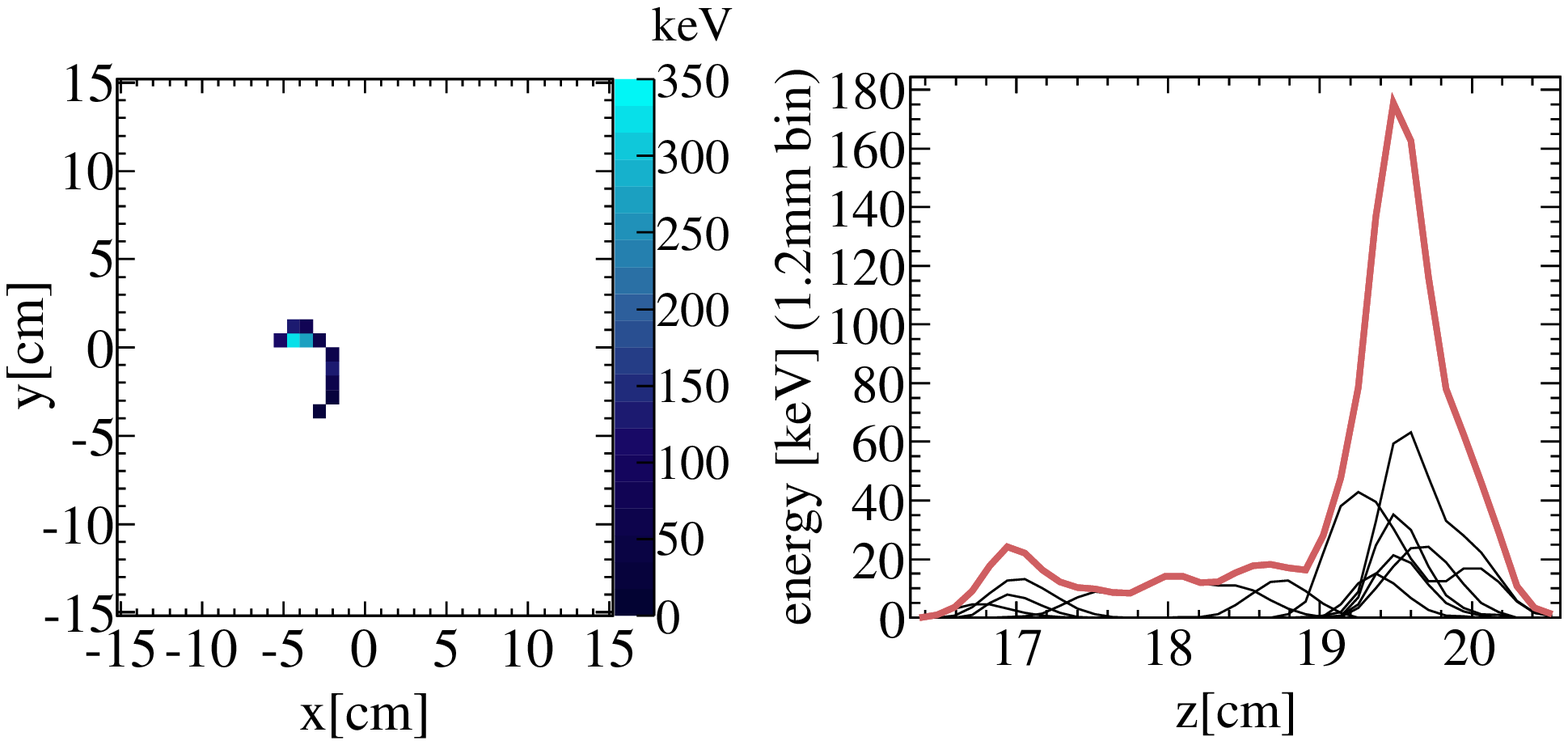}
 \includegraphics*[width=7.1cm]{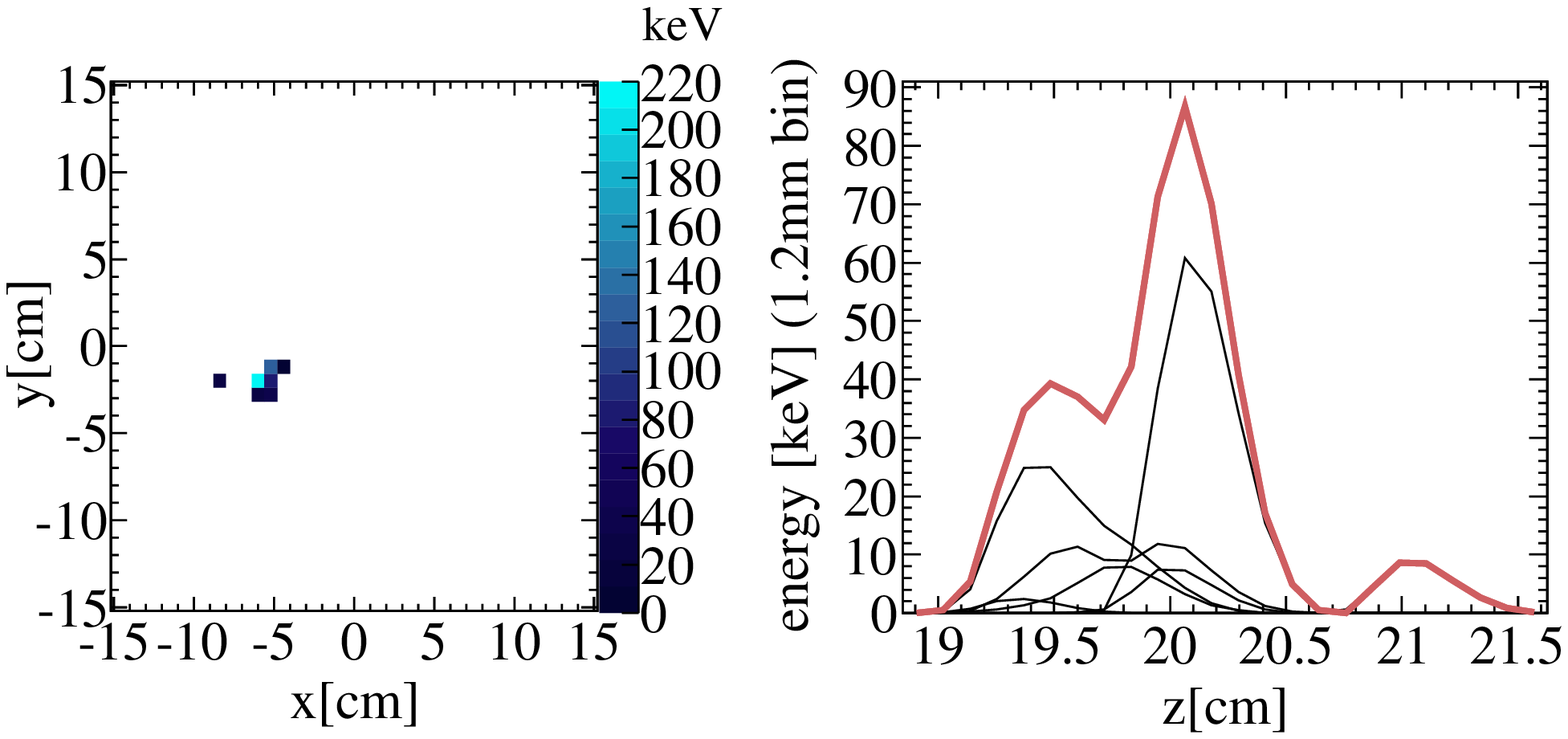}
 \includegraphics*[width=7.1cm]{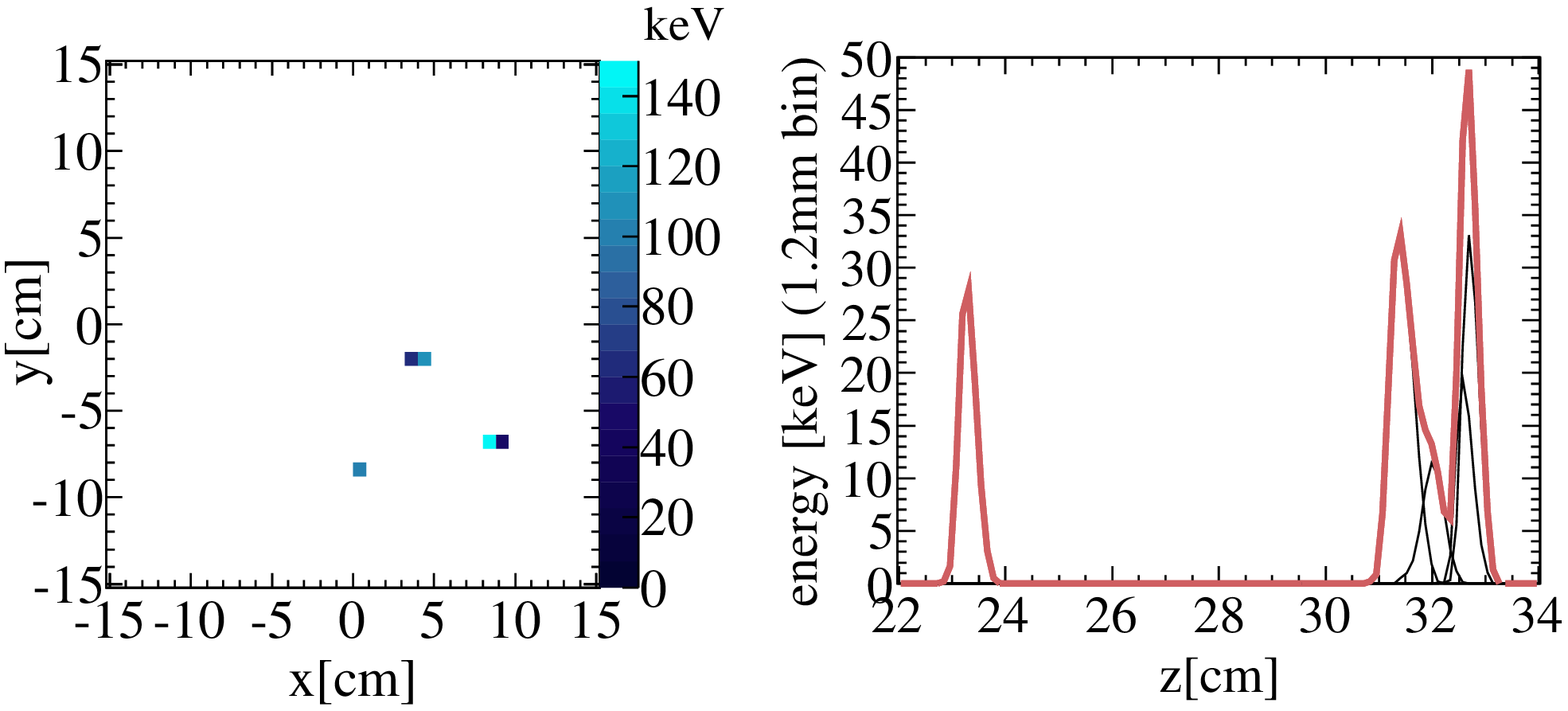}
 \caption{Several events obtained after calibration: $x$-$y$ projection (left) and $z$-projection (right). Top panel: 1.275 MeV photo-electron. Middle panel: 0.511 MeV photo-electron and characteristic X-ray emission. Bottom panel: 0.511 MeV multi-Compton event. On the right column the thin lines represent the individual pulses registered for each pixel, and the thick line represents the event pulse (sum of all pulses).}
 \label{PulseExample}
 \end{figure}

The data analysis proceeds starting from the 511-point waveform recorded by each of the $288\times4$ readout channels, with an event being a collection of waveforms. The analysis chain resembles, although supersedes, the one in \cite{DiegoTMA}:

\begin{enumerate}
\item\label{z-p} {\it Zero-suppression}. A global threshold is chosen relative to the system equivalent noise charge (ENC) at about $\varepsilon_{th}=5$-$10\times$ENC, set to minimize the presence of noisy channels and cross-talk: a) the non-firing pixels and b) the non-firing time bins of the fired pixels are removed from the analysis chain. The threshold is defined over a certain charge pedestal, that is determined pulse-by-pulse and subtracted.
\item\label{PSA} {\it Basic pulse shape analysis}. Basic parameters of each waveform being above threshold are obtained: $x$-$y$ pixel coordinates, position along the drift region ($z$), waveform integral ($\varepsilon$), width ($\sigma_t$), etc.
\item {\it Cluster finder}. Two dimensional ionization clusters (projected in the $x$-$y$ plane) are identified and tagged with a recursive depth-first search (DFS) method. For operational purposes a cluster is considered to be connected if a path exists between any two pixels belonging to it; a path is a superposition of segments; a segment is a line joining two neighbours; each pixel has 8 neighbours. Examples of events consisting of 1, 2, and 3 clusters are given in Fig. \ref{PulseExample}.
\item {\it Identification of displaced X-rays}. Displaced clusters with an energy in the range $20$-$40$ keV are identified and undergo a Gaussian fit to refine their main parameters relative to the estimate in (\ref{PSA}).
\item \label{cal} {\it Quality cut}. Events with fluctuating base-lines, anomalous noise or saturation are discarded.
\item \label{fid}{\it Volume fiducialization}. Events not fully contained within 1 cm of the boundaries of the active region $[\phi, 0$-$h]$ are removed.
\item \label{cut}{\it Suppression of random coincidences}. Random coincidences are suppressed by resorting to the physical correlation between a waveform's width $\sigma_t$ and its position along the drift coordinate $z$. This correlation, stemming from the longitudinal electron diffusion, is particularly strong for X-ray clusters due their nearly point-like nature (see Fig. \ref{SigmaVsZ} for a graphical explanation).
\item {\it Event selection and analysis}. Physical criteria are imposed in order to select the events of interest.
\end{enumerate}

In the following, a magnitude $A$ may be attributed to a generic event ($A_{evt}$), to a gamma-ray event ($A_{\gamma}$), to an electron track within an event ($A_e$) or to an X-ray within an event ($A_x$): $A \equiv x,y$ refers to energy-weighted positions projected into the readout plane, $A \equiv z$ refers to the arrival time of the maximum of the associated sum-pulse, $A \equiv \sigma$ to its width, $A \equiv \varepsilon$ to its sum-energy.

 \begin{figure}[h!!!]
 \centering
 \includegraphics*[width=7.5cm]{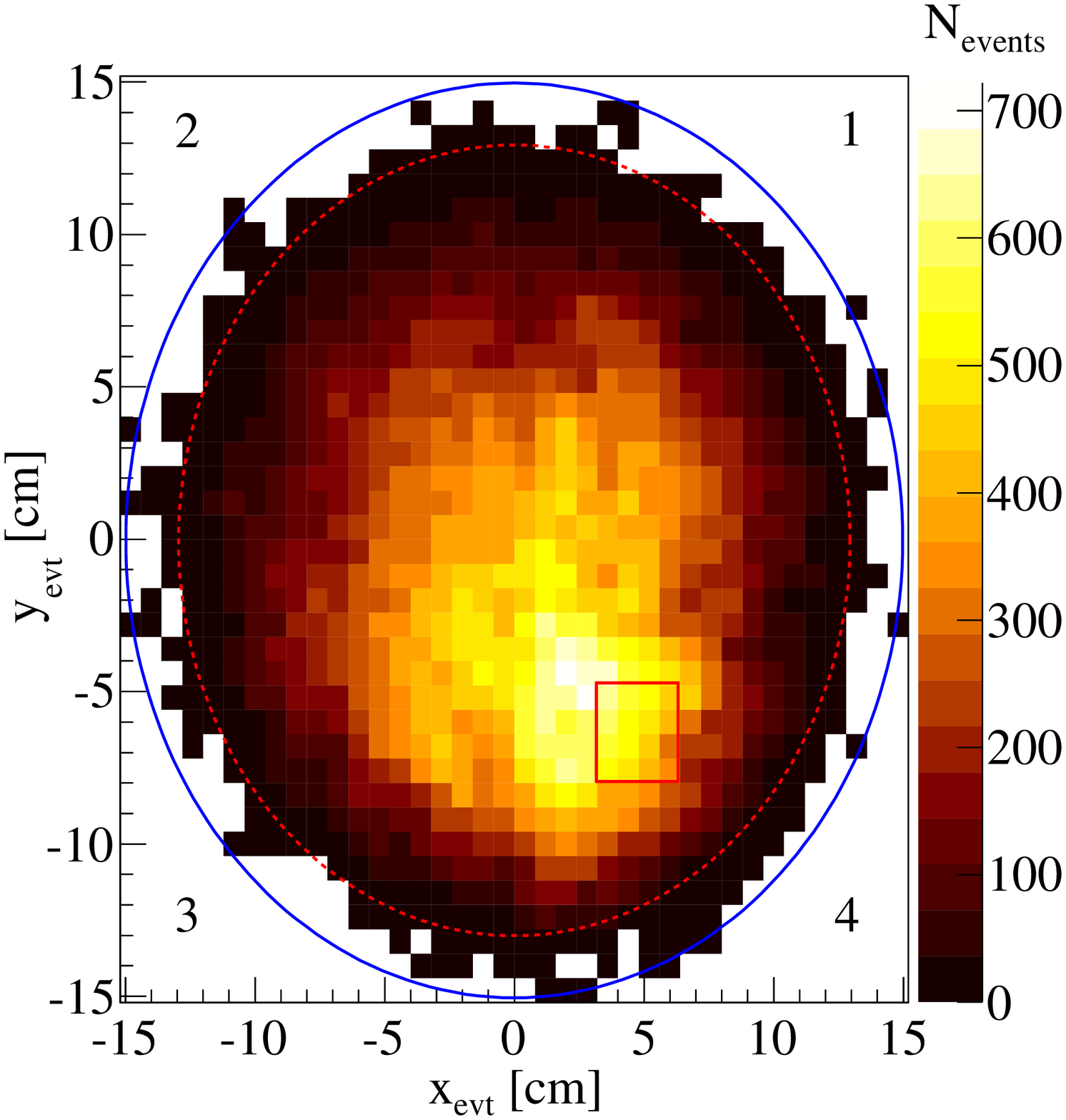}
 \includegraphics*[width=8.9cm]{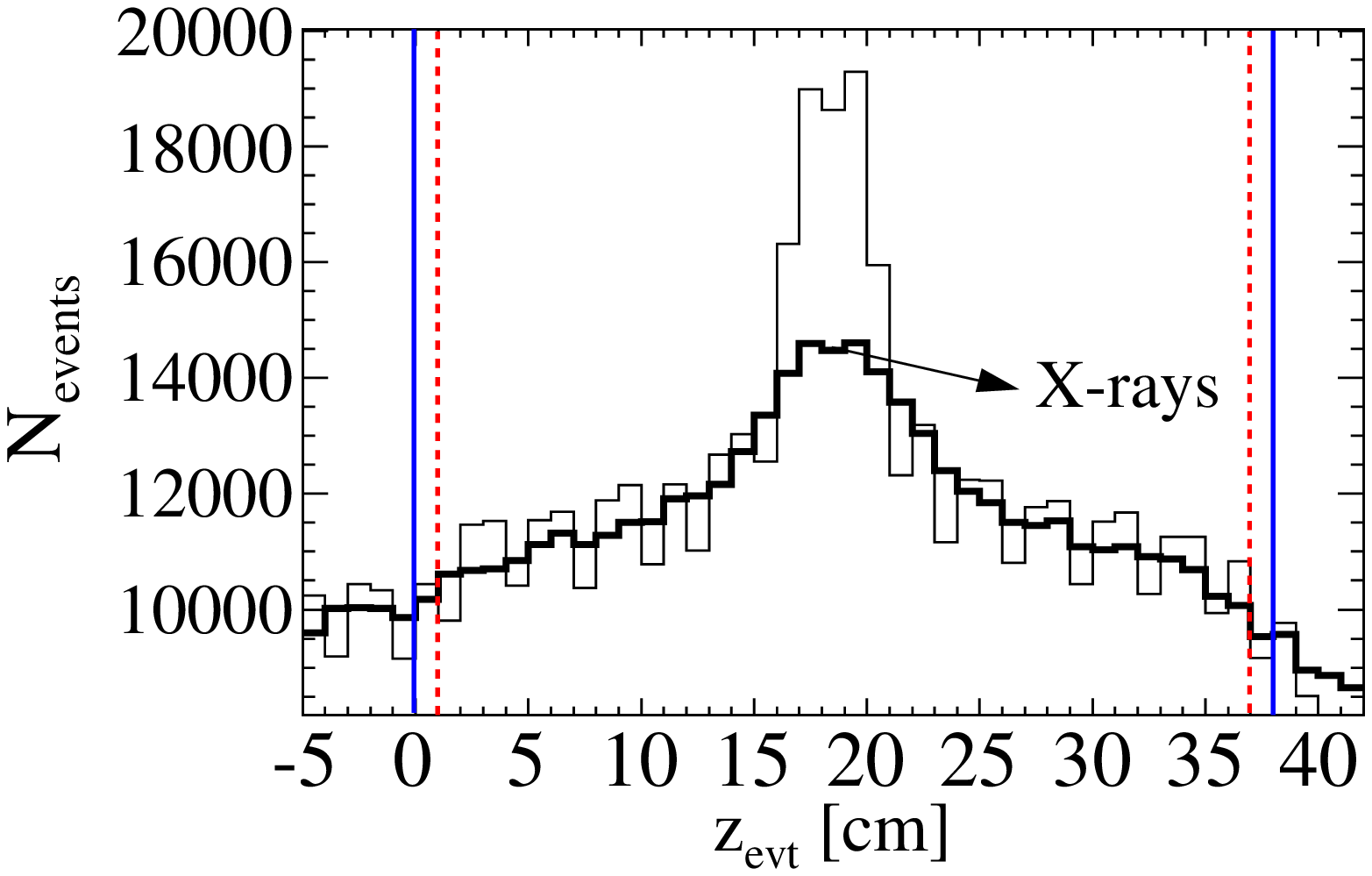}
 \caption{Top panel: event position (energy-weighted) on the readout plane for a typical $^{22}$Na run at 10 bar (with source placed at about the lower right corner, 5~cm distance from the readout boundaries). Micromegas readout sectors appear numbered according to their geometrical position. Bottom panel: $z$ event position (at the waveform's maximum) of a sub-sample of events displaying characteristic X-ray emission (thin line). The reconstructed position of the X-rays, smeared due to their finite mean free path, is shown with the thick line. The physical boundaries of the readout plane and drift region are indicated with continuous lines, while dashed lines indicate the fiducial volume.}
 \label{XY2D}
 \end{figure}

\subsection{Calibration}

Data are first examined after step (\ref{cal}), Fig. \ref{XY2D}. Reference positions are identified in the distribution of arrival times of the events' ionization, providing an estimate of the $z$ position (Fig. \ref{XY2D}-down). The slope of the $\sigma_t$ vs $z$ correlation for X-ray clusters is stored for future cuts. Gain variations from sector to sector are corrected (10-20\%) and the integral of the waveforms converted to energy.

After these preliminary calibrations, a pixel-by-pixel gain equalization is performed iteratively by aligning the pixel energy distributions obtained over the $x$-$y$ plane. Such pixel distributions are formed by histogramming the energy of X-ray clusters for which the contribution of the given pixel is the highest. This calibration procedure, that involves a mild sector-wise correction of gain transients (5-10\%), converges after 2-5 iterations when the energy resolution reaches an optimum.

 \begin{figure}[ht!!!]
 \centering
 \includegraphics*[width=4.3 cm]{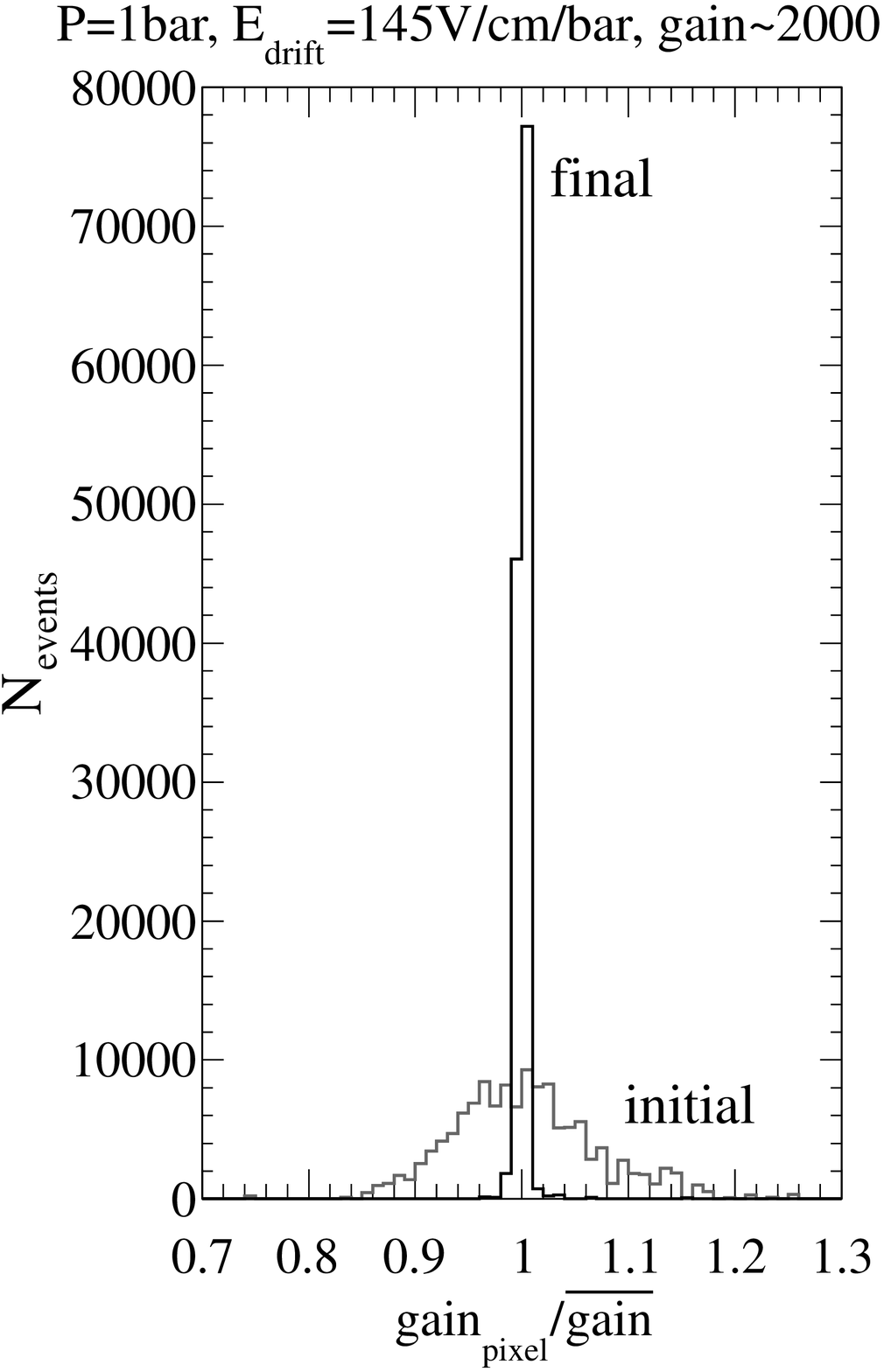}
 \includegraphics*[width=4.3 cm]{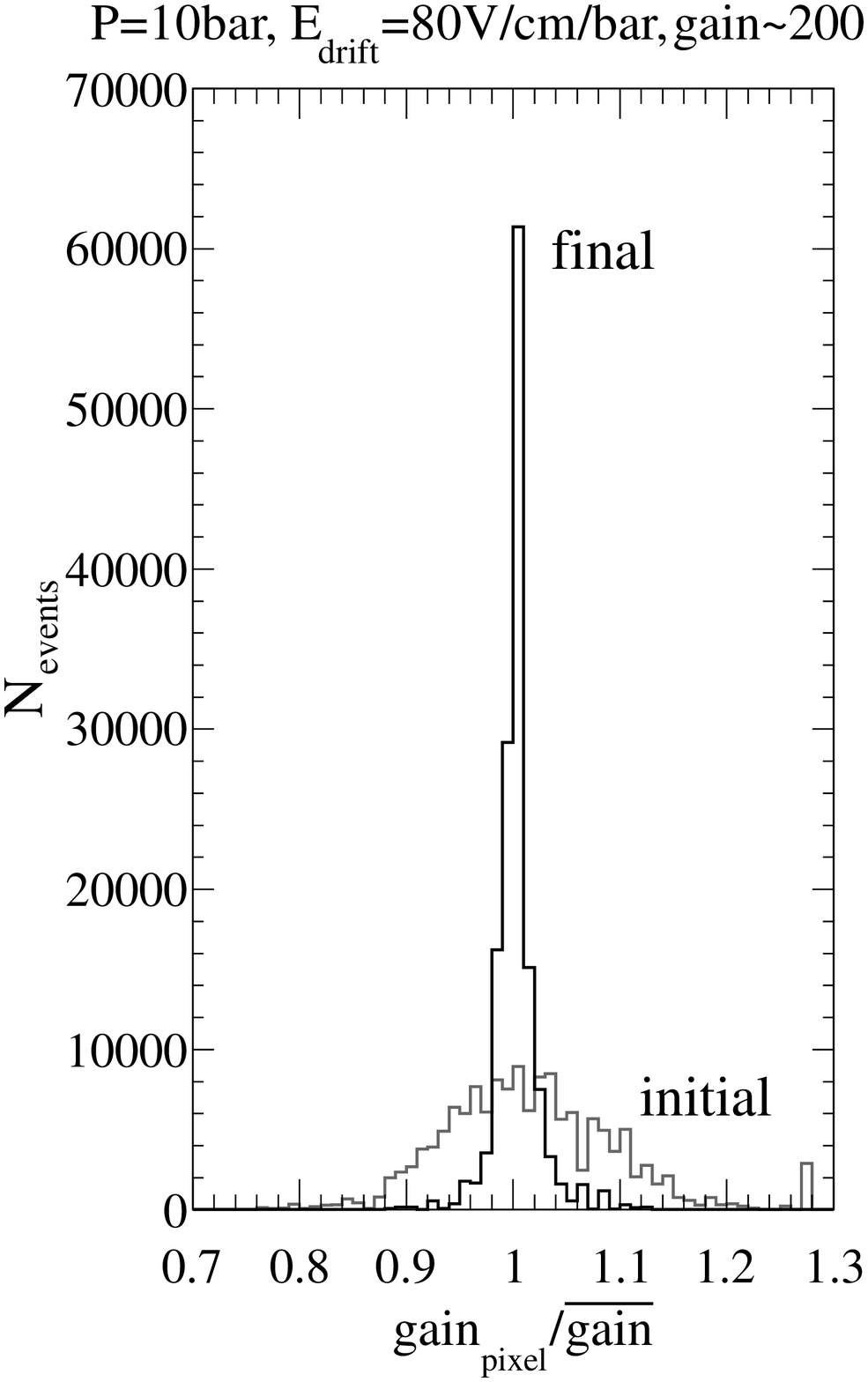}

 \includegraphics*[width=4.3 cm]{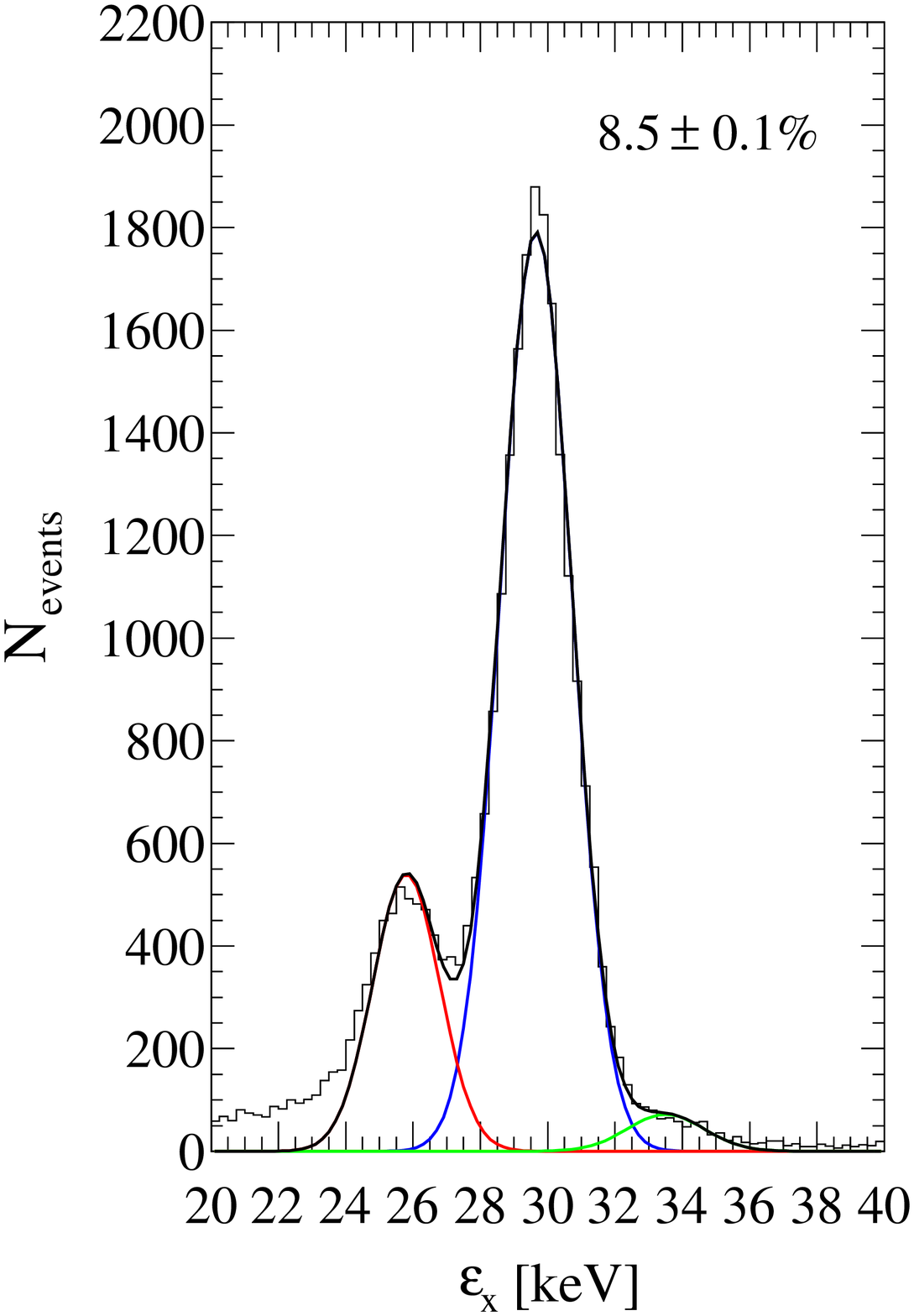}
 \includegraphics*[width=4.3 cm]{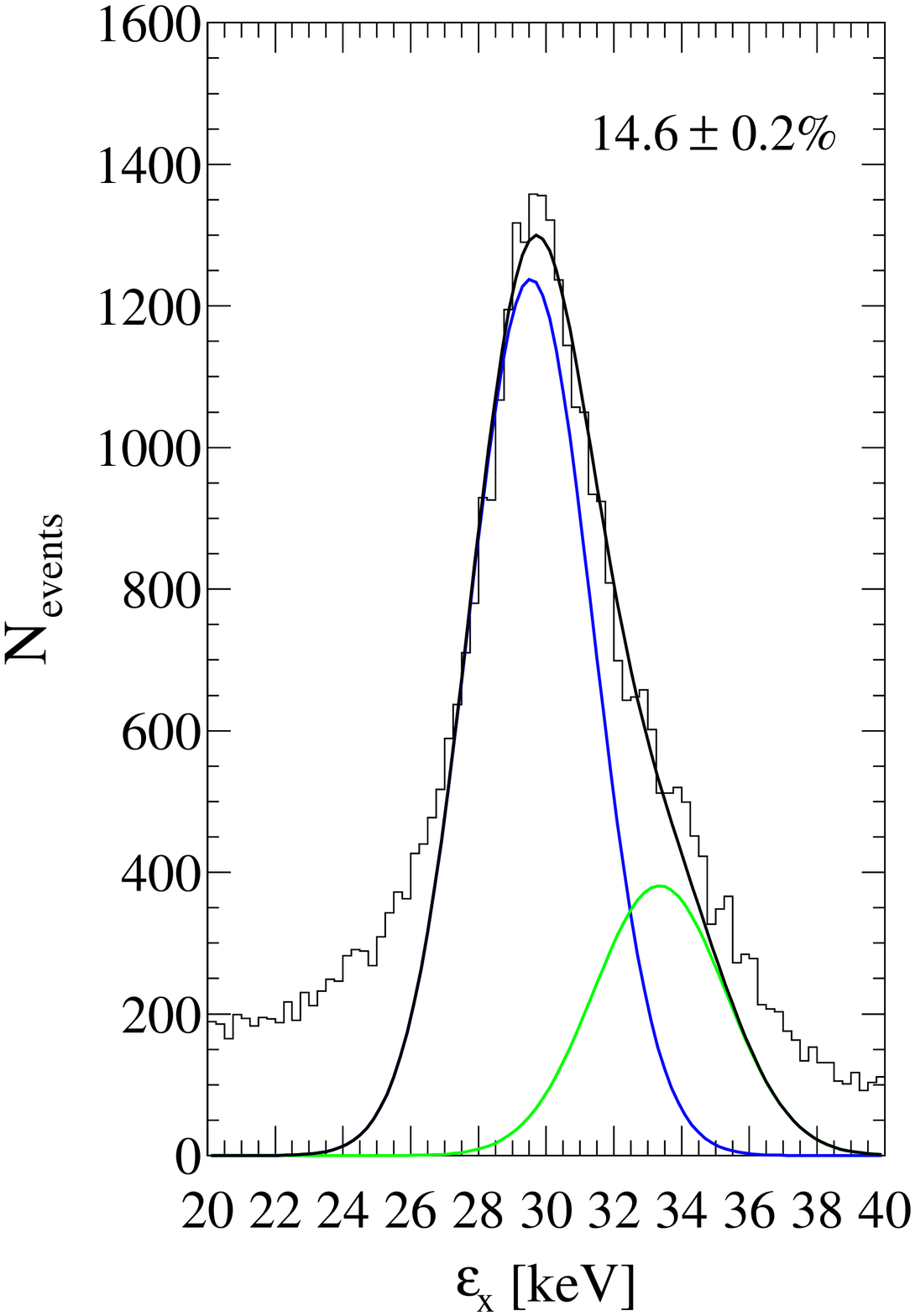}

 \caption{Top panel: histogram showing the relative pixel gain (weighted by the number of events per pixel) before and after calibration, for 1~bar (left) and 10~bar (right). Bottom panel: system resolution (FWHM) after calibration, suppression of random coincidences and TPC fiducialization for 1~bar (left) and 10~bar (right). Only the $K_{\alpha}, K_{\beta}$ lines and escape peaks are considered in the fit, with normalization free. The spectrum to the left is dominated by the escape peaks from the $^{241}$Am 59.5 keV emission, while right shows characteristic Xe emission from the $^{22}$Na 0.511, 1.275 MeV interactions.}
 \label{ResDist}
 \end{figure}
The distribution of the per pixel ratios of the $\sim\!30$ keV X-ray peak energy relative to the average value over the whole readout plane
are shown in Fig. \ref{ResDist}-up, each entry being weighted by the number of X-rays counted in that pixel.
These distributions, obtained before and after the calibration procedure, inform about the robustness and quality of the algorithm, their final widths being below 1\%(1 bar) and 2\%(10 bar) FWHM, nowhere near influencing the system resolution obtained after calibration: $8.5\pm0.1\%$(1 bar) and $14.6\pm0.2\%$(10 bar), Fig. \ref{ResDist}-down. No sizeable deterioration of the latter values could be appreciated upon reduction of the statistics available for calibration by a factor $\times 4$.

\begin{figure}[ht!!!]
 \centering
 \includegraphics*[width=9cm]{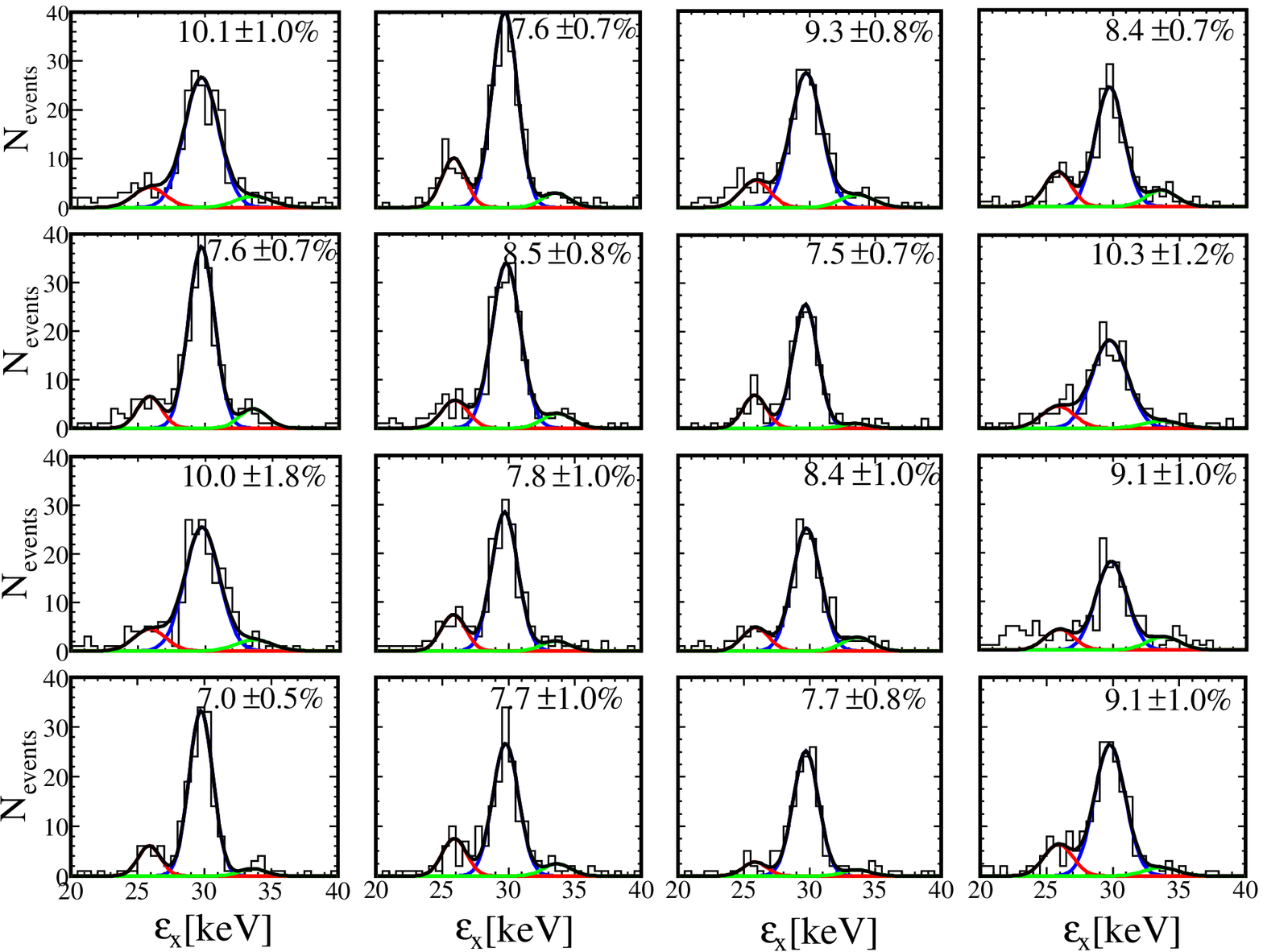}
  \includegraphics*[width=9cm]{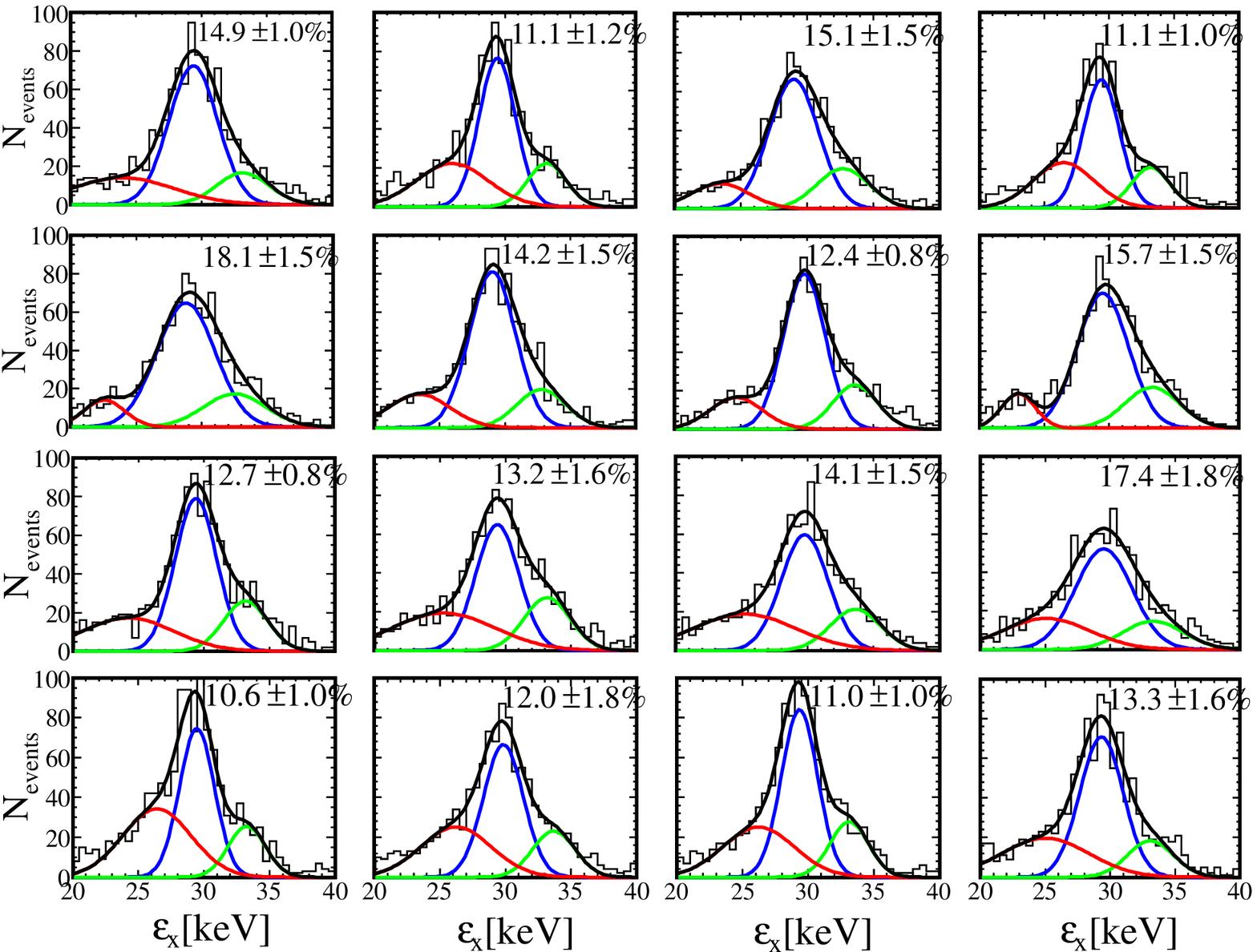}
  \caption{X-ray energy distributions after calibration, obtained on a selected $4\tn{pixel} \times 4\tn{pixel}$ region. Spectra are described through an iterative multi-Gaussian fit consisting of $K_{\alpha}$, $K_{\beta}$ characteristic Xenon emission and escape peaks. Peak positions and widths are bound relative to the $K_{\alpha}$ position by the known energy differences and $1/\sqrt{\varepsilon}$ scaling, with normalization left free. In case of bad fit quality or low statistics, a simple iterative gaussian fit is performed. Top panel: 1 bar. Bottom panel: 10 bar. At high pressure only characteristic $K_{\alpha}$, $K_{\beta}$ emission is expected, therefore the mean, width and normalization of the (unphysical) leftmost Gaussian are left free. This roughly accounts for the charge losses (due to a relatively high threshold around 5~keV) and increases the stability of the fit.}
 \label{CalDist1bar}
 \end{figure}

Fig. \ref{CalDist1bar} shows the energy distributions and fitting functions used for calibration in a $4\tn{pixel} \times 4\tn{pixel}$ region (squared area in Fig. \ref{XY2D}), after calibration converged. A description of the fit details has been captioned. Although a peak-alignment with the anticipated $\sim\!\!1$\% precision can be appreciated, there is a significant variability in the width of the main peak, beyond the statistical uncertainty of the fit.

\subsection{Long-term stability and technical performance}

At 10~bar, about 11 million $\gamma$-events from a 15~kBq $^{22}$Na source were taken during April-July 2014, totalling 100 live-days. System operation at nominal working conditions proceeded continuously for 24 h a day and largely without manual intervention: no dedicated shifts were necessary except for sporadic monitoring of pressure and the current readings of the HV power supply for safety reasons. The bias of the readout plane was controlled with a N1471 CAEN power supply, that would halt operation sector-wise in the event of an over-current exceeding 200~nA for more than 10~s. Effective data takings took place for about 30\% of the total live-time, with the interval between acquisition runs dedicated to analysis and diagnostics.

The level of pixel functionality was 90\% throughout the measurements, the non-functionality approximately shared between permanently damaged pixels and connectivity issues (the latter discussed elsewhere \cite{comm}). Most of the damage arose during earlier measurements with $\alpha$-sources and low-energy X-rays in \cite{DiegoTMA, comm} as a consequence of sustained over-current leading to violent sparking, the main causes identified being: i) readout mishandling, ii) $\alpha$-emitter contamination of the cathode after prolonged use of a Rn source (\cite{comm}), and iii) damage during normal operation with $\gamma$-sources at relatively high gains. In case of damage, a manual intervention rendered the pixel electrically floating, restoring the electrostatic equilibrium in the affected region. The fringe fields resulting from this indirect biasing of the pixel potential do not penetrate beyond 1\% of the neighboring pixels, thus ensuring normal detector operation except for the modest geometrical loss associated to the intervention.

\begin{figure}[ht!!!]
\centering
\includegraphics*[width=9cm]{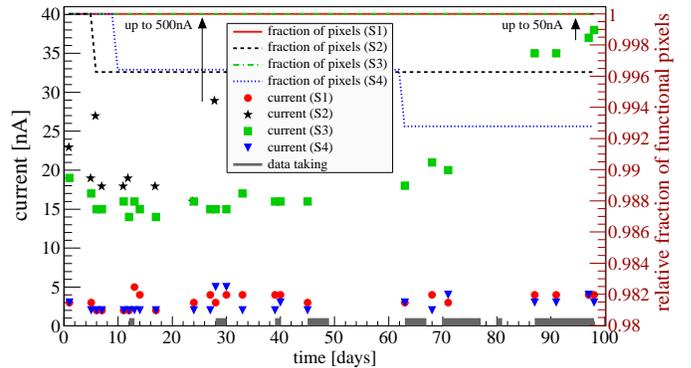}
\caption{Evolution of the dark current and relative fraction of functional pixels in the Micromegas readout for 100 live-days under continuous exposure of the TPC to a 15 kBq $^{22}$Na source. The normalization is arbitrarily set to 1 at time=0, factoring out the damage occurring during commissioning (5\%) and the presence of unconnected pixels (see text for details). Three pixels were damaged during this prolonged operation (sector 2 and 4). Grey rectangles show the periods when data takings took place.}
\label{currents}
\end{figure}

Although dark currents normally stayed within 0.015-0.03 nA/cm$^2$, values as high as 0.3 and even 3 nA/cm$^2$ were reached in two of the sectors (Fig. \ref{currents}). This did not affect detector operation except for a somewhat $\times 1/2$ lower gain in the most extreme case. The absence of a strong gain drop or hot spot suggests that these leaks may be related to loss of insulation downstream the amplification gap. Weak points may appear at connectors and cable traces that result indirectly biased when a pixel is left floating during the aforementioned process.

The presence of micro-discharges in non-insulated Micromegas operated close to breakdown is a well documented phenomenon (see for instance \cite{DiegoRev} and references therein), resulting in the present case in an average rate of current excursions about 10 /min/m$^2$. Provided the pre-amplifier input is protected they have no influence in the detector stability. As shown below, they have also a negligible impact in the system exposure, since their natural scale is that of few amplification holes. In most cases this phenomenon affects only one pixel ($C_{pix}=20$ pF) during a short period of time. More intense current excursions are exponentially less frequent although potentially more harmful.

Four types of discharges can be found in this system, qualitatively enumerated below in order of increasing strength:

\begin{enumerate}[I]
\item Low intensity: current excursions at the level of 20-60 nA.
\item Medium intensity: current excursions exceeding the HV current-limit of 200 nA, resulting in a HV trip.
\item \label{HVtrip}  High intensity: current excursions exceeding the HV current-limit of 200 nA during 10 s, resulting in a HV ramp down.
\item \label{deadPix} Permanent damage: a high intensity discharge that permanently develops a high conductivity path.
\end{enumerate}

Incidentally, stronger micro-discharges were found to have also a stronger impact on the fraction of system exposure that is affected, $\frac{Mt|_{aff}}{Mt}$. This magnitude can be approximated by the micro-discharge rate, and the area ($A_{aff}$) and time ($\Delta T_{aff}$) affected after each occurrence, as detailed in Table \ref{discharges}:

\beq
\frac{Mt|_{aff}}{Mt} \simeq rate \Delta A_{aff} \Delta T_{aff}
\eeq

After accounting for all instabilities in the readout plane, the overall reduction in exposure is at the level of $0.5 \times 10^{-3}$, seemingly of little practical importance. Moreover, if an automatic method would be implemented for leaving a damaged pixel in open-circuit after \ref{deadPix}-type discharges, operation without manual intervention becomes possible in m$^2$ experiments for Micromegas operating close to breakdown.

\begin{table}[h]
  \centering
  \begin{tabular}{|c|c|c|c|c|c|}
     \hline
     type & rate & $\Delta$T$_{aff}$ & $\Delta$A$_{aff}$ & $\Delta$A$_{loss}$ & Mt$|_{aff}$/Mt \\
     \hline
     I   & 10/(min m$^2$) & 2 s    & pixel   &     -       & $0.02 \times 10^{-3}$ \\
     II  & 2/(h m$^2$)    & 10 s   & sector  &     -       & $0.09 \times 10^{-3}$ \\
     III & 5/(d m$^2$)    & 2 min  & sector  &     -       & $0.14 \times 10^{-3}$ \\
     IV  & 1\%/y          & 2 h    & sector  & pixel & $0.20 \times 10^{-3}$ \\
     \hline
   \end{tabular}
  \caption{Types of micro-discharges in the Micromegas readout together with the (approximate) affected time, affected area and damage, if any. In case IV the time is merely indicative since it involves manual intervention. The last column shows the fraction of experiment's exposure affected, amounting in total to less than 0.05\%.}\label{discharges}
\end{table}


A priori more worrisome for long-term operation is the initial damage and the rate of new damage. As for the initial 5\% damage, it is difficult to state which part is unavoidable during the commissioning and system conditioning. Certainly 1\% levels would be approachable through a factor $\times 2$ reduction in pixel size and improved system handling. Implementation of a tight quality control (QC) would bring additional gain. The chosen TPC orientation maximizes the accumulation of dust, that should be avoided in an experiment if no particular geometrical constraint exists. The remaining damage rate of 1\%/year would benefit in a similar way from the above measures.

\section{Response to $30$ keV X-rays and comparison with earlier results}\label{SwarmSec}

\subsection{Energy resolution and intrinsic performance} \label{Eres}

\begin{figure}[ht!!!]
 \centering
 \includegraphics*[width=9cm]{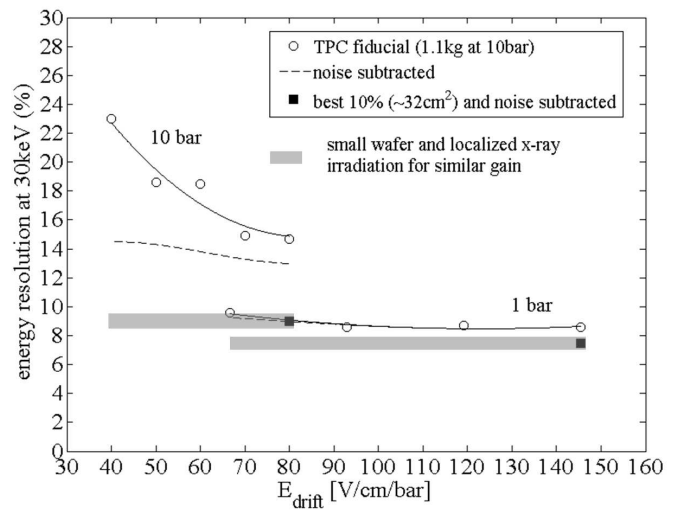}
 \caption{Measured energy resolution (FWHM) as a function of the drift field in the TPC fiducial region (circles), together with trend lines. Noise-subtracted values are given by the dashed line, while squares indicate the result after both pixel selection and noise subtraction, for the highest fields. Shadowed bands represent the values obtained earlier in small setups and for similar gains in \cite{XeTMA}.}
 \label{ResSummary}
 \end{figure}

The energy resolution obtained for X-rays after calibration is compiled as a function of the drift field in Fig. \ref{ResSummary} (circles). The horizontal bands show, as a reference, the energy resolution obtained in small setups and for similar gains \cite{XeTMA}. We refer to this resolution, hereafter, as `intrinsic' (to the charge-amplification readout). Since such an intrinsic response offers a promising extrapolation to the $Q_{\beta\beta}$ energy we discuss below the observed excess.

On the one hand the degradation with lower drift field cannot be attributed to attachment or recombination since charge losses from 80 to 40~V/cm/bar amount to some 10\% (subsection \ref{swarmSec}), hence implying a negligible contribution to the resolution below 3\% (to be summed in quadrature).\footnote{This follows the canonical formula for the case where recombination $\mathcal{R}$ is decoupled from other processes: $2.35\sqrt{\mathcal{R}/(1-\mathcal{R})~ n_o}$, $n_o=\varepsilon/W_I \simeq 1200 e^-$ for 30~keV X-rays in Xenon.} Instead, the reason for the observed behaviour can be found on a relatively low S/N ratio and a reduced sampling frequency at low drift fields. The latter is determined by the limited number of sampling points provided by the electronics (511) in connection with the reduced drift velocity. In fact, given the response function of the AFTER chip for $\delta$-excitation:

\beq
f(t)=A e^{-3t/\tau} (t/\tau)^3 \sin(t/\tau) \label{response}
\eeq
even for the longest shaping time available ($\tau=2 \mu$s), the minimum sampling time achievable (if recording the complete drift region) barely satisfies the Shannon-Nyquist condition for the lowest drift fields.
Additional contributions to this response such as the ion transit time along the Micromegas amplification gap or the extension of the electron track are expected to be small and indeed $\tau=2.1 \mu$s described well the temporal width of X-rays found close to the anode.

\begin{figure}[ht!!!]
 \centering
 \includegraphics*[width=7.3cm]{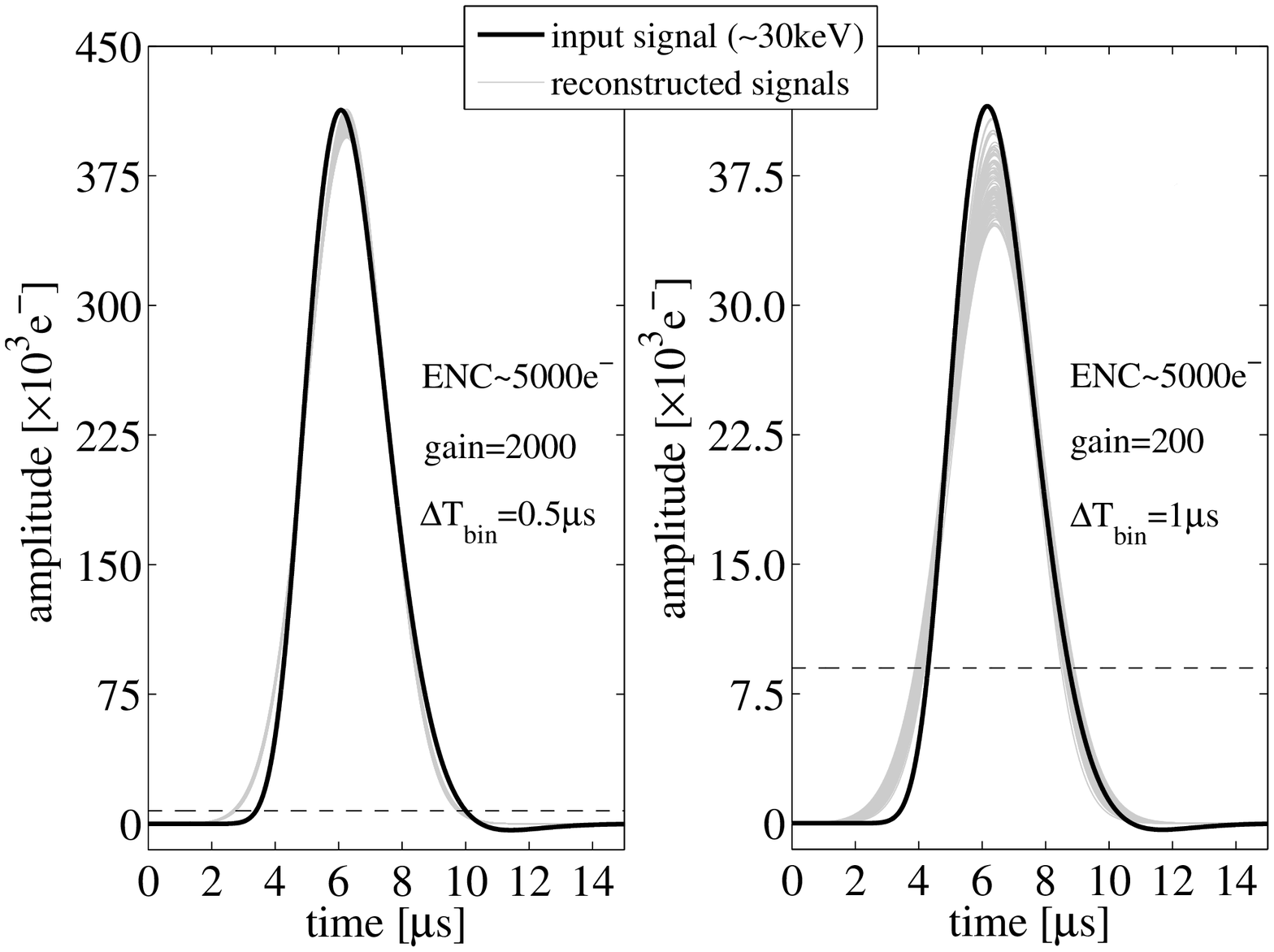}

 \includegraphics*[width=7.3cm]{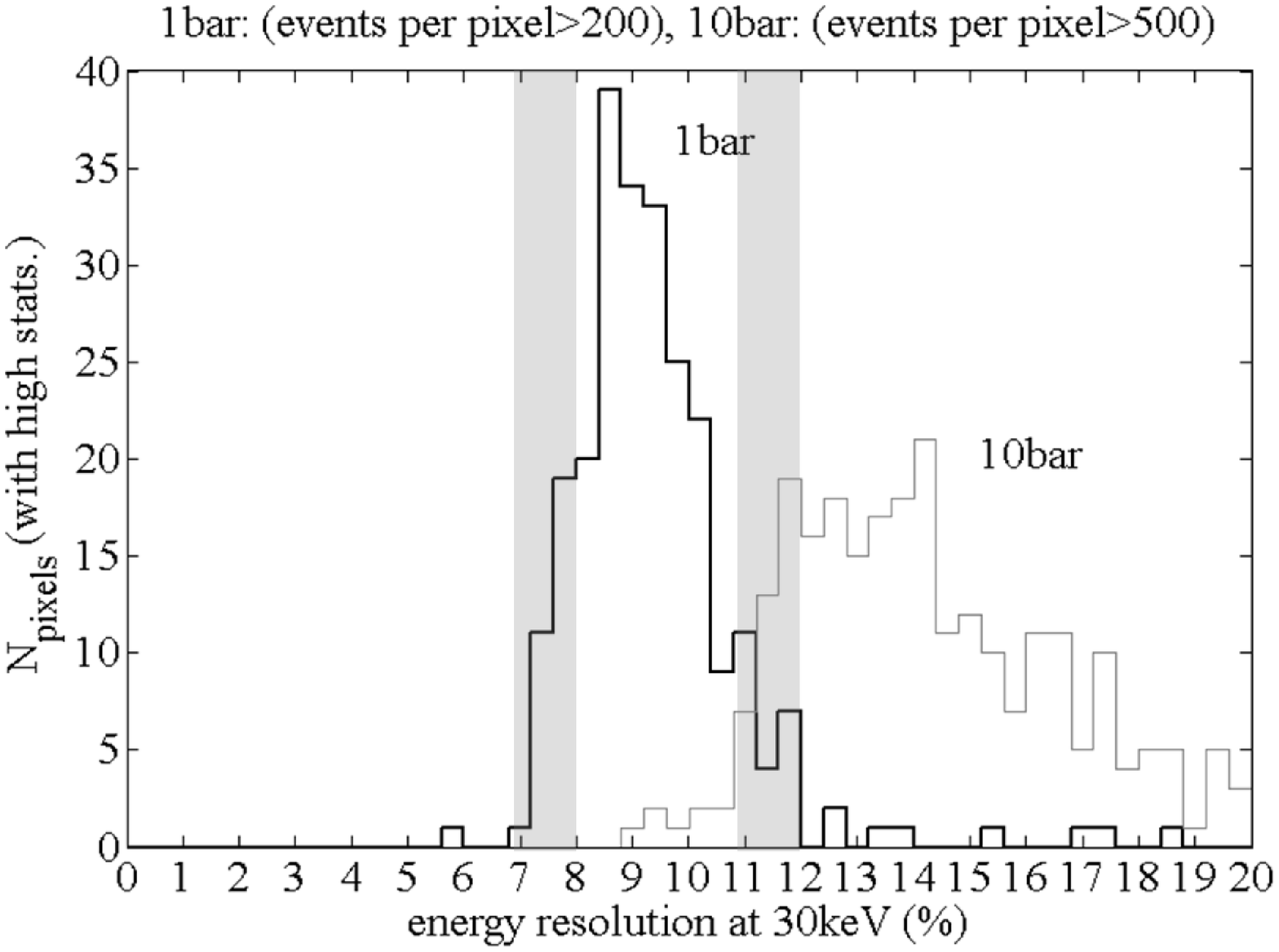}
 \caption{Contributions to the energy resolution beyond the intrinsic expectation from the Micromegas readout. Top panel: simulation of the reconstruction algorithm (grey lines) for the assumed response function given in black at the nominal thresholds (dashed). Left and right illustrate typical running conditions for low and high pressure, respectively. Bottom panel: distribution of the pixels' energy resolutions (FWHM) for 1 and 10 bar. Shadowed bands correspond to an approximate selection of the best 10\% after removing outliers.}
 \label{ResDistVert}
 \end{figure}

From the above perspective it can be expected that the deterioration of the energy resolution at low fields comes from the inability to accurately reconstruct the integral of $f(t)$, as illustrated by simulation in Fig. \ref{ResDistVert}-up. For clarity, $f(t)$ displays here the amplitude of a 30~keV charge deposit after including the longitudinal charge spread at about mid-drift (black line), as seen by the acquisition system. The resulting waveform is randomly sampled with the corresponding time bin $\Delta T_{bin}$ and smeared with white noise, a Gaussian fit finally implemented exactly as done in the data analysis (grey lines). Simulations indicate that, if assuming ENC$=7500$e$^-$ (50\% in excess of the average ENC measured over the whole readout plane), most of the observed deterioration in the reconstructed pulse integral can be explained (dashed lines in Fig. \ref{ResSummary}).


A second source of deterioration can be found in the aforementioned variation of the energy resolution from pixel to pixel, illustrated in Fig. \ref{ResDistVert}-down by histogramming the energy resolution of those pixels with higher statistics. Even if accepting the presence of statistical and systematic effects in the fits of Fig. \ref{CalDist1bar}, some additional contribution seems to be needed to explain the spread of these distributions. In particular, the larger spread observed at high pressure can be qualitatively understood from the steeper behaviour of the gain vs amplification field ($E_{a}$) systematics in \cite{XeTMA}, with the quantity $\frac{\Delta gain}{gain}/\frac{\Delta E_{a}}{E_{a}}$ increasing with operating pressure. A recent microscopic simulation of the Micromegas readout shows that the energy resolution at high pressure is indeed much more vulnerable to the geometrical tolerances in the diameter of the amplification holes, by a factor $\times 3$ \cite{RefElisa}. If approximately selecting the best 10\% pixels (outliers removed) values closer to intrinsic can be found (shadowed bands in Fig. \ref{ResDistVert}). An approximate decomposition of the energy resolution for the highest drift fields studied is:

\bear
\!\!\!\!&&\mathfrak{R}_{1bar} \!\!=\! \sqrt{ 0.075^2(\tn{int})\!\!+\!0.002^2(\tn{S/N})\!\!+\!0.035^2(\tn{p-p})}\\
\!\!\!\!&&\mathfrak{R}_{10bar}\!\!=\! \sqrt{ 0.09^2(\tn{int})\!+\!0.07^2(\tn{S/N})\!+\!0.09^2(\tn{p-p})}
\eear
that includes the contributions intrinsic to the readout (int), signal to noise (S/N) and pixel-to-pixel (p-p).
It must be noted that collimation at the level of few mm$^2$ (much smaller than the pixel size in NEXT-MM) is often used in test setups and that there is an implicit selection bias in that under-performing wafers are discarded before characterization or suffer from early damage. Microbulk Micromegas even on modest sizes of 10's of cm$^2$ as those used in CAST already show energy resolutions mildly higher than intrinsic \cite{CASTres}. Put simply, any residual gain variation below the minimum resolvable unit of a pixel (or a strip, in that case) is hardly correctable during an experiment.

\subsection{Parameters of the electron swarm}\label{swarmSec}

The precision in the topological reconstruction of the ionization trail left by a primary track in a TPC depends on the native system voxelization (pixel size, time bin) and the `swarm' properties of the electrons released: drift velocity $v_d$ and pressure-normalized diffusion coefficients $D_{L,T}^*$. The reconstructed energy may be limited, on the other hand, by losses due to electron-ion recombination or electron attachment ($\mathcal{R}$, $\eta$):

\bear
& \frac{Q_E(z)}{Q_\infty(0)} & = \mathcal{R} e^{-\eta z} \label{attach} \\
& \sigma_{z,xy}(z) & = D_{L,T}^*\frac{\sqrt{z}}{\sqrt{P}} \\
& \sigma_{t}(z) & = \frac{\sigma_{z}}{v_d} \label{sigmaT}
\eear
$Q_E(z)$ refers to the charge collected as a function of the electric field and drift distance, and $Q_\infty(0)$ represents the amount of charge released by the event in the absence of recombination.
These electron swarm parameters can be extracted from the same X-ray sub-sample used for the calorimetric analysis of subsection \ref{Eres}, provided their point-like nature allows approaching the $\delta$-type ionization conditions assumed in eqs. \ref{attach}-\ref{sigmaT}: i) the drift velocity is determined from the difference in arrival times of X-ray clusters produced at two well defined positions and ii) the longitudinal diffusion $D_L^*$ and attachment coefficients are obtained from the $\sigma_t^2$ vs $z$ and $\varepsilon$ vs $z$ cluster behaviour (eqs. \ref{attach}, \ref{sigmaT}).
It is in principle possible to extract the transverse diffusion coefficient $D_{T}^*$ through a separate analysis, however the spatial sensitivity at 10~bar was found not to be sufficient for this. Details on the extraction of $D_{L,T}^*$, $v_d$ for 1-3 bar can be found in \cite{XeTMA} and so here only the 10 bar analysis is described.
\begin{figure}[ht!!!]
\centering
\includegraphics*[width=8.0cm]{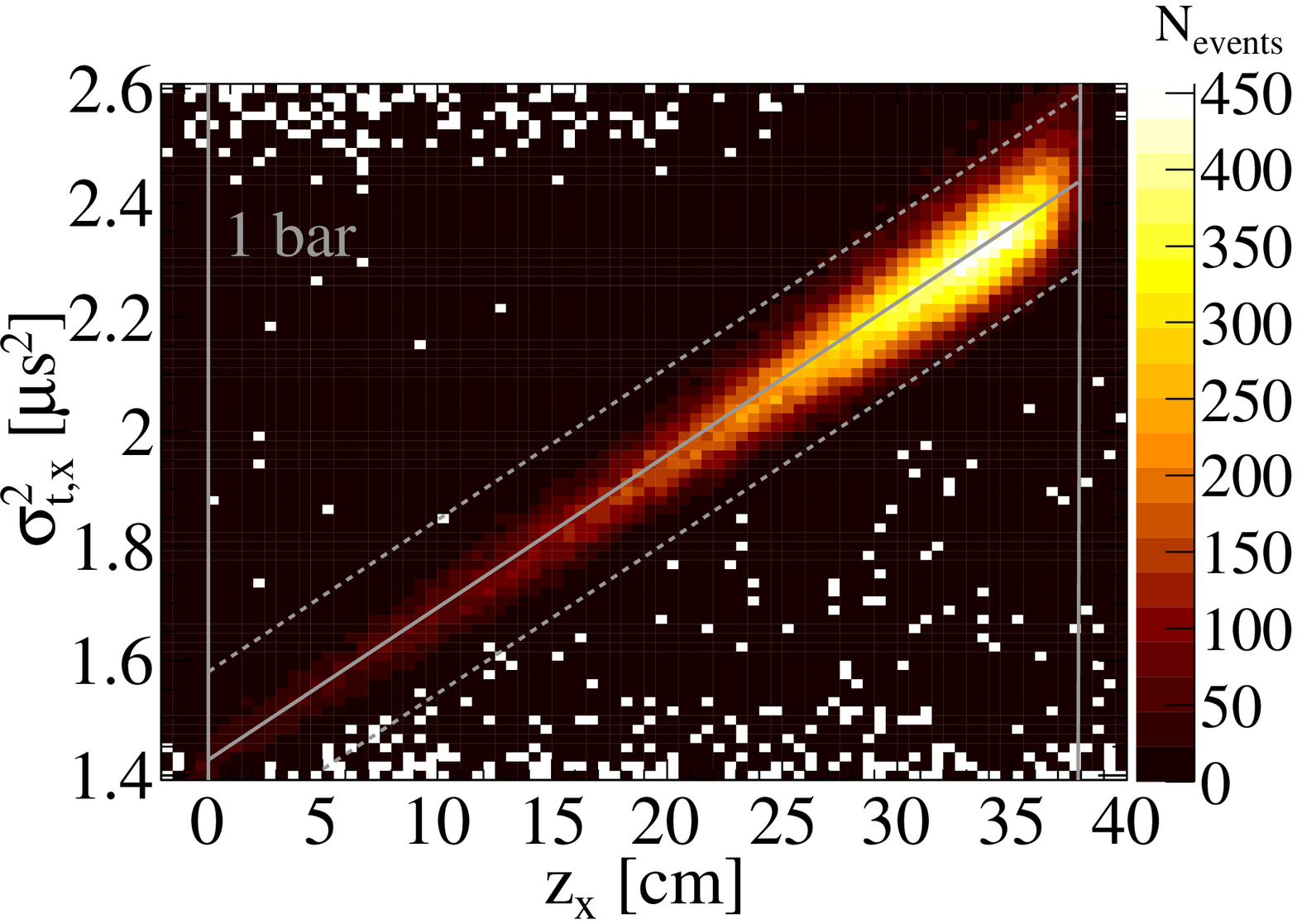}
\includegraphics*[width=8.0cm]{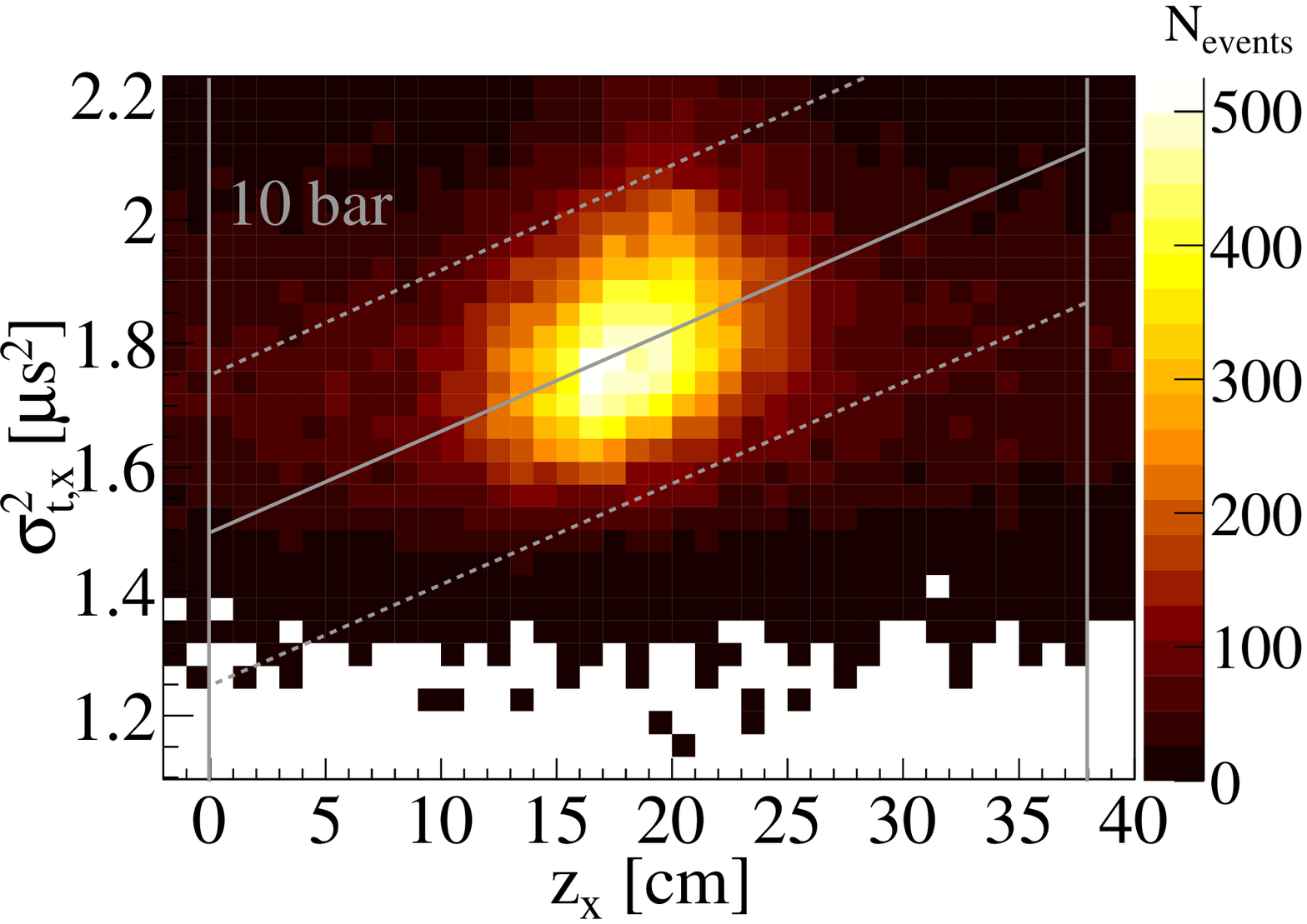}
\caption{Temporal variance of the pulse as a function of the drift position for isolated X-rays. The boundaries of the drift region as well as the trends are indicated by continuous lines. Dashed lines together with the drift boundaries delimit the cut region selected for suppressing random coincidences. Top panel: 1 bar (source placed above cathode). Bottom panel: 10 bar (source placed at mid-drift).}
\label{SigmaVsZ}
\end{figure}

The drift velocity was determined from the TPC anode and $^{22}$Na source positions, the latter identified as a pronounced peak in the X-ray time distribution, corresponding to $z=19\pm2$ cm (Fig. \ref{XY2D}). Arrival times from clusters in the TPC-anode region are synchronous with the NaI(Tl) trigger signal and could be identified with an estimated 2 cm uncertainty.

\begin{figure}[ht!!!]
\centering
\includegraphics*[width=9.0cm]{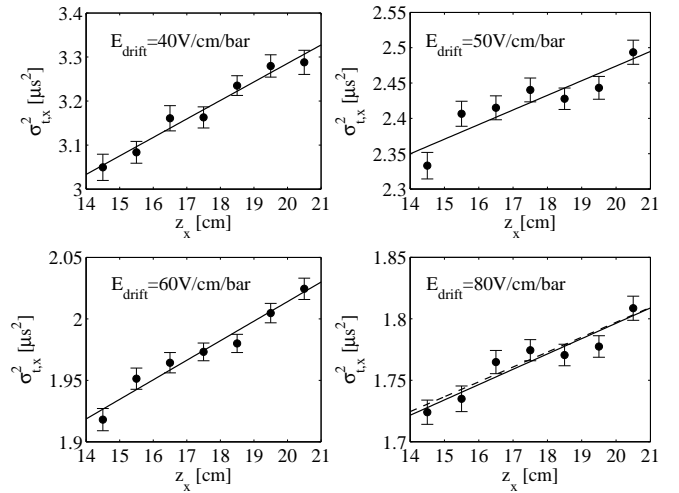}
\caption{Determination of the slope from the $\sigma_t^2$ vs $z$ plots for X-ray clusters. An iterative Gaussian fit has been performed in 1 cm steps and the slope obtained from a weighted fit. The plot at 80 V/cm/bar includes the results from simulation (dashed lines).}
\label{SigmaVsZfit}
\end{figure}

Concerning longitudinal diffusion, the $\sigma_t^2$ vs $z$ correlation considerably widened at 10 bar as compared to 1 bar (Fig. \ref{SigmaVsZ}). This is consistent with the 10-fold S/N reduction and the associated deterioration in the reconstruction of the pulse characteristics.
In order to minimize the additional blurring introduced by random coincidences at 10 bar, the main electron track associated to the X-ray was requested to be in the range $z_{e}=19\pm5$ cm.
Linear fits in a selected region in $z$ around the source position yielded the desired slopes (shown in Fig. \ref{SigmaVsZfit}). For illustration, the simulation of the pulse reconstruction procedure at various $z$-positions was performed, and the final lineal fit is shown for the 80 V/cm/bar case (dashed line). The parameters used in the simulation are $\tau=2.15$ $\mu$s, ENC=7500e$^-$, $\Delta{T}_{bin}=1$ $\mu$s, $\varepsilon_{th}=10\times$ENC, gain$=200$, and the slope introduced was the measured one. It is important to note that the systematic upwards shift (resulting from the effective increase of the pulse width during reconstruction) is correctly captured in simulation, and it is qualitatively reproduced for lower fields where the effect is considerably larger. However, even in this highly non-ideal situation, the reconstructed slope stays within 5\% of the value introduced, at most. The uncertainty in the determination of the longitudinal diffusion coefficient is obtained from this 5\%, the 10\% uncertainty in the drift velocity, the statistical error of the linear fit and the variations for different assumptions on the fitting range. When summed in quadrature, the resulting uncertainty lies in the range 10-15\%.

\begin{figure}[ht!!!]
\centering
\includegraphics*[width=9.1cm]{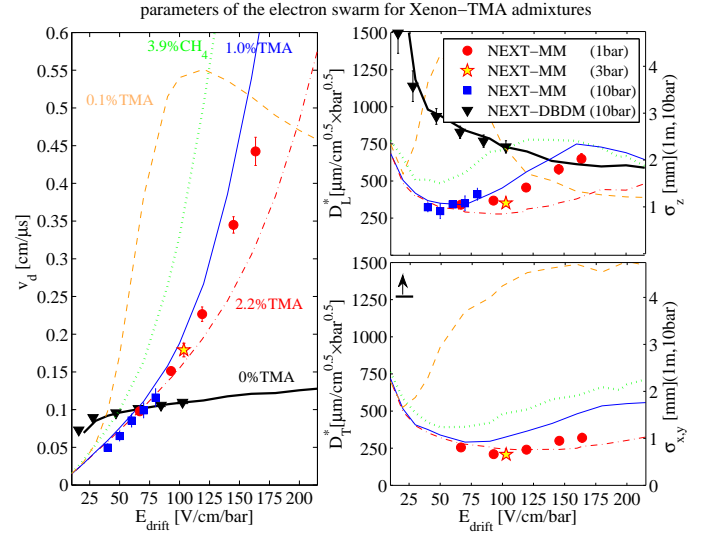}
\caption{Compilation of the main properties of high pressure Xenon TPCs for 1(1.0) 3(2.7) and 10(10.1) bar and different TMA admixtures, together with simulation (Magboltz 10.0.1, \cite{Magboltz}), as a function of the reduced electric field. Left: drift velocities. Right: longitudinal and transverse diffusion coefficients. Simulation results for Xe-CH$_4$ (Gotthard) and Xe-TMA in a 0.1\% admixture are shown for illustration.}
\label{Swarm}
\end{figure}

Fig. \ref{Swarm} shows a compilation of the main results in the field region that is of interest for the NEXT experiment together with a comparison with the microscopic code Magboltz 10.0.1 \cite{Magboltz}. Additional measurements performed in small setups at various pressures can be found in \cite{DiegoTMA} (drift velocity) and \cite{HerreraReco} (drift velocity and longitudinal diffusion coefficient), showing good mutual agreement. The combined results convey evidence of the strong electron cooling properties of TMA, showing a reduction in diffusion within factors of $\times 3$(L), $\times 20$(T) of the ones achievable in pure Xenon, and significantly outperforming the expectation for Xe-CH$_4$ admixtures \cite{Gotthard} by nearly a factor $\times 2$. Xe-TMA sets the diffusion scale in Xenon mixtures to 1 mm for 1 m drift at 10 bar, less than a part in 200 relative to the size of a $Q_{\beta\beta}$ event at the same pressure.

Interestingly, the cross-over shown by the drift velocity in Xe-TMA mixtures relative to pure Xenon is well reproduced, and so below 75~V/cm/bar the mixture becomes actually slower than pure Xenon. Since this type of systems often operate in a closed gas loop, a high sensitivity of the drift velocity to the reduced field may turn to be a relevant practical advantage, allowing the optimization of the voxelization along $z$ during operation.

\begin{figure}[ht!!!]
\centering
\includegraphics*[width=8.1cm]{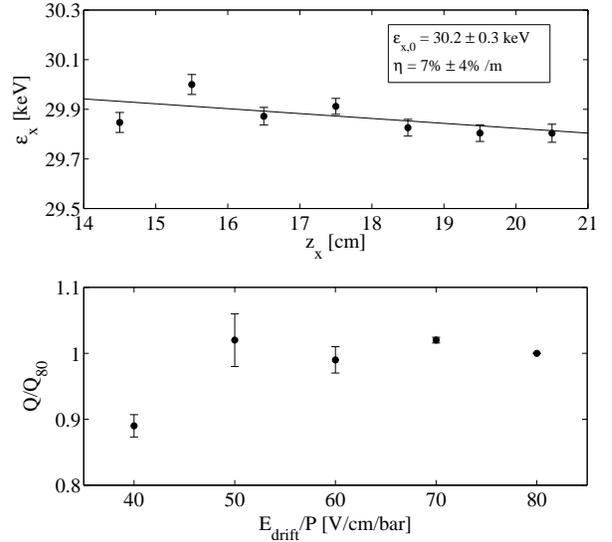}
\caption{Top panel: X-ray energy peak position as a function of drift distance for $E_{drift}=80$ V/cm/bar. Bottom panel: X-ray energy peak position (back-converted to charge) relative to 80 V/cm/bar after correcting for attachment (only statistical errors shown).}
\label{Attach}
\end{figure}

The dependence of the X-ray energy peak with the drift distance is shown in Fig. \ref{Attach}-up for $E_{drift}=80$ V/cm/bar. Further to the $z_{e}$ cut, random coincidences have been suppressed through a $\sigma_t^2$-$z$ correlation cut (dashed lines in Fig. \ref{SigmaVsZ}). The resulting $\varepsilon$ vs $z$ distribution is binned and fitted iteratively to a Gaussian function, yielding for the highest field an attachment coefficient $\eta=7\%\pm 4\%$/m after an exponential fit (corresponding to an electron lifetime of 12~ms).

At last, the relative fraction of primary ionization $Q$ that is recombined can be obtained by studying the variations of the X-ray energy peak position (back-converted to charge), obtained at about mid-drift and corrected for attachment. Due to the anticipated smallness of the effect the analysis has several caveats, in particular in view of the variation of the sampling time and diffusion with field. Simulation indicates that any systematic shift of the reconstructed pulse integral is below 1\% even for the lowest fields, and diffusion is relatively stable (Fig. \ref{Swarm}), so we may neglect any effect related to the pulse shape reconstruction. The remaining systematic effects are connected to the presence of (not understood) gain transients at the level of 5\% and to the stronger $z$-dependence of the reconstructed energy at low fields, implying up to a 10\% correction. Under these assumptions a coarse estimate of the maximum recombination observed in present X-ray data is $\mathcal{R}_x=1-Q_{40}(0)/Q_{80}(0)=0.12\pm0.02(\tn{sta}) ^{+0.15}_{-0.05}(\tn{sys})$, the statistical uncertainty coming from sector-to-sector variations (Fig. \ref{Attach}-down). Within such a large uncertainty, this value is compatible with the charge loss reported in \cite{RefElisa}, that sets a scale for recombination at the level $\mathcal{R}_x = 1-Q_{80}(0)/Q_{\infty}(0) \! \lesssim 0.20$ in $1\%$ TMA admixtures at 10~bar.

\section{Extended electron tracks from $\gamma$-rays}\label{ExtTracks}

\begin{figure}[ht!!!]
\centering
\includegraphics*[width=8cm]{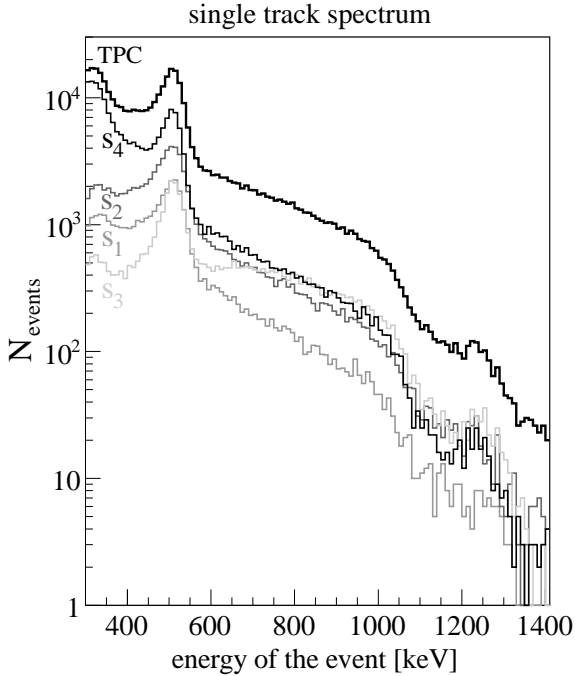}
\caption{Single-track energy reconstruction in the TPC after calibration (thick line), for events coming from a $^{22}$Na source in coincidence with a Na(Tl) detector. The contributions from tracks fully contained in each of the individual sectors ($S_{1,2,3,4}$) are also shown.}
\label{Spectrum}
\end{figure}

A typical $\gamma$-ray coincidence spectrum obtained for single-tracks after calibration is shown in Fig. \ref{Spectrum}. The TPC working conditions are: Xenon/trimethylamine filling gas at 99/1 volume admixture, $P=10.1$ bar ($T=20$ deg), $E_{drift}=80$ V/cm/bar, $E_{amp}=11.1~$kV/cm/bar. These settings correspond to a gain for the Micromegas readout of about $\times200$, with the main TPC charge-drift properties being $v_d=1.16 \pm 0.12 $ mm/$\mu$s, $\sigma_{_L}(1$m)$=1.3\pm 0.13$ mm, $\sigma_{_T}(1$m)$=0.8\pm 0.15$ mm (from Fig. \ref{Swarm}), $\eta=7\% \pm 4 \%$/m (Fig. \ref{Attach}). Indirect estimates for initial charge recombination, Penning transfer probability and primary scintillation yield are $\mathcal{R}_{x}\lesssim0.2$ (section \ref{swarmSec}), $r_P\geq0.1$ and $S_1(\sim\!\!300\tn{nm})\geq50$ ph./MeV (given later in section \ref{PF}), respectively. The remaining parameters impacting the accuracy of the reconstructed track are the threshold energy $\varepsilon_{th}=4$ keV (ENC=5000e$^-$), the response function (dominated by the AFTER chip) of 1.2 $\mu$s-$\sigma$ (1.4 mm) and a time sampling of 1 $\mu$s (1.2 mm), that ultimately results in a native 3D-voxelization of 8 mm$ \times 8$ mm$ \times 1.2$ mm.

The most prominent features in the spectrum correspond to the photo-peaks from the annihilation of the e$^+$ emitted in the $^{22}$Na decay, at 0.511~MeV, and the $\gamma$-emission after the daughter's decay at 1.275 MeV. The energy resolution is not sufficient to separate the escape peak associated to the 0.511~MeV interactions, appearing as a shoulder to the left of the main peak. For sector $2$ (where the gain was lower) and sector $1$ (that showed larger gain transients) the tails to the left are more pronounced and the resolution worsened. The suppression of the 0.511~MeV region in sector $3$ relative to the high energy Compton contribution is due to the trigger geometry (the solid angle for 0.511 MeV is determined by the coincidence with the NaI(Tl) detector, covering mostly sectors 1, 2 and 4).

\subsection{Calorimetry}

We focused on events that satisfy the analysis criteria \ref{cal}-\ref{fid} (subsection \ref{analysis}), and concentrate in particular on electron tracks displaying characteristic X-ray emission.
This additional criterion allows suppressing the escape peaks and multi-Compton events as well as random coincidences. The latter can be achieved through the $\sigma_t^2$ vs $z$ correlation cut on the X-ray cluster.\footnote{It can be noted that, coincidentally, the selected sub-sample is the one used for extracting the electron swarm parameters, where emphasis was put on the X-ray behavior.}

An example of a typical 0.511~MeV event is shown in Fig. \ref{PulseExample} (middle panel).
Electrons with this energy deposit about 1/2 of their energy in the track's last stage as a result of the Bethe-Bloch increase (Bragg peak) and
the increased multiple scattering, a region generically referred to as `blob'. The approximately equal energy sharing
between the blob and the (nearly) `mip' region results in relatively small tracks of 3-4 cm length. As later shown,
electrons from the 1.275 MeV region can reach, on the other hand, considerably longer extensions of up to 10-15 cm (Fig. \ref{1200}).


\begin{figure}[ht!!!]
\centering
\includegraphics*[width=4.3cm]{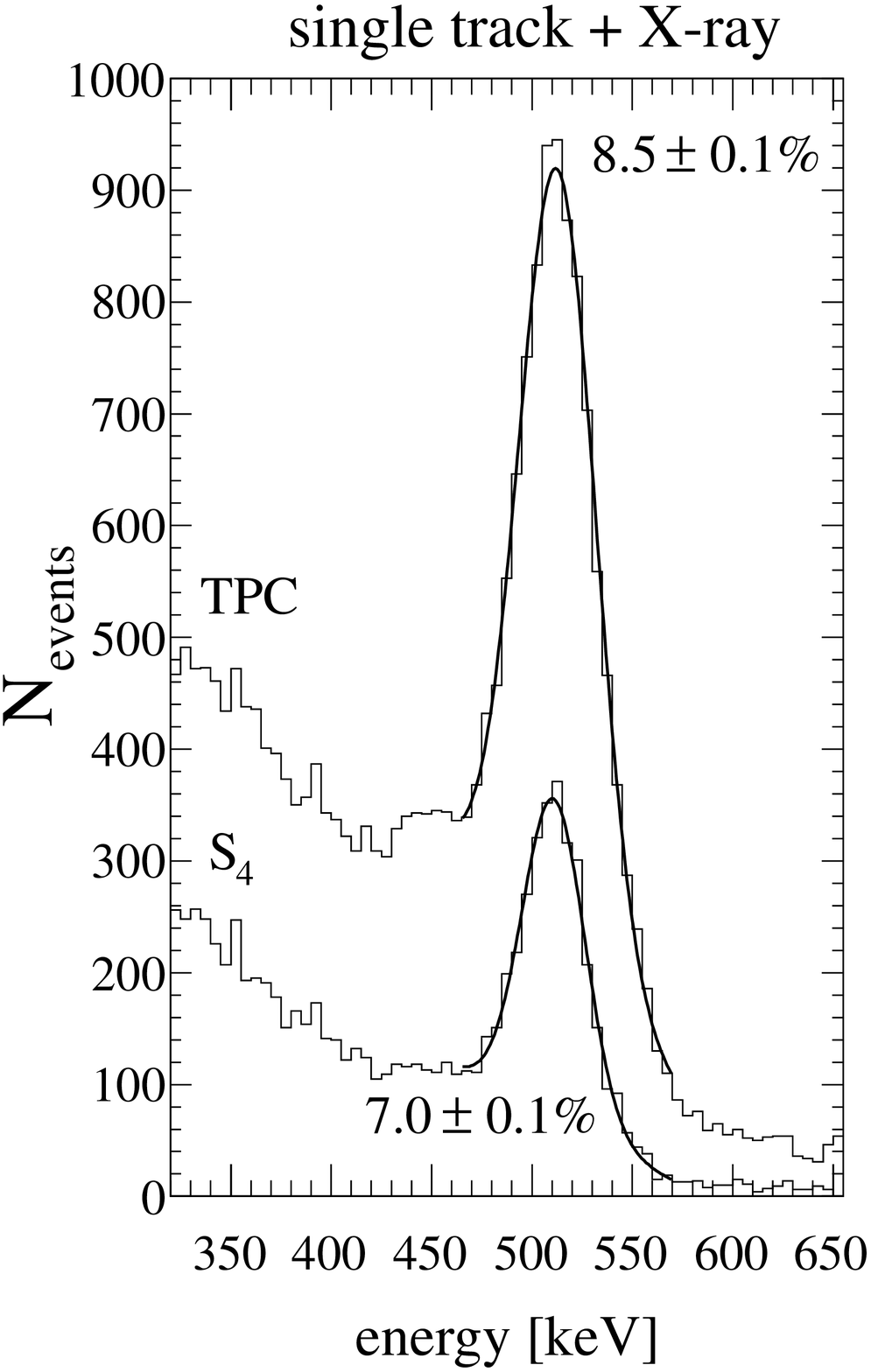}
\includegraphics*[width=4.3cm]{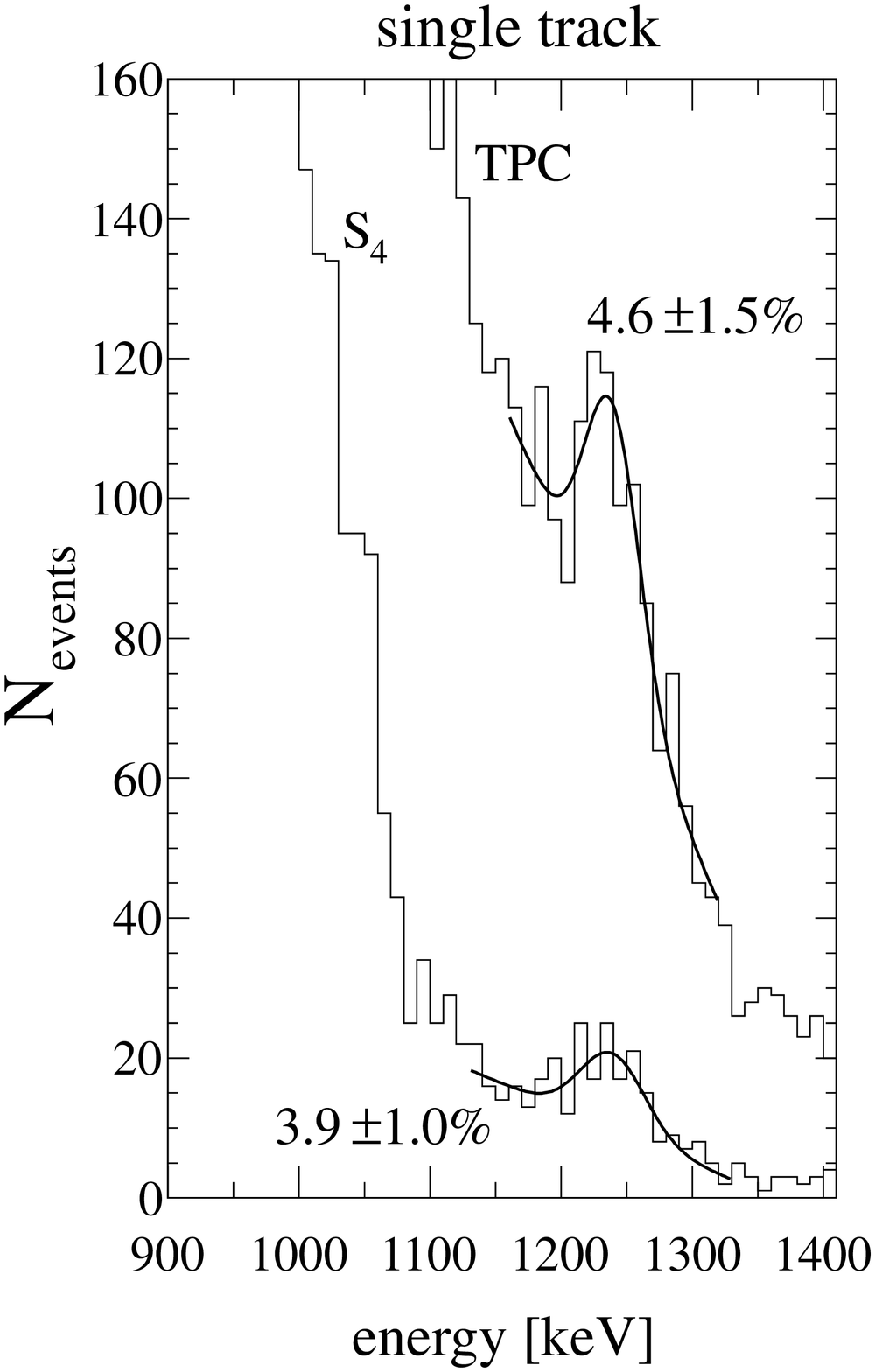}
\caption{Left: energy distribution around 0.511 MeV obtained for double-tracks (i.e., events consisting of a single-track in the presence of a displaced X-ray), shown for the TPC fiducial region and for sector 4. Right: energy distribution around 1.275 MeV for single-tracks.}
\label{EnerResGammas}
\end{figure}

The spectrum obtained in the 0.511 MeV region for the full TPC after fiducialization (1.1~kg) is consistent with an energy resolution of $8.5\pm0.1\%$ (Fig. \ref{EnerResGammas}-left), while the sector with the highest statistics and S/N (S$_4$) performs at $7.0\pm0.1$\%.
The energy resolution obtained in the 1.275 MeV region can be estimated from the high statistics single-track sample as 4.6$\pm1.5$\%(TPC) and 3.9$\pm1.0$\%(S$_4$), Fig. \ref{EnerResGammas}-right. Therefore we come to energy resolutions for MeV-electron tracks of $4.5$-$5.5\% \!\sqrt{1 \tn{MeV}/{\varepsilon}}$ (0.511~MeV) and $4.4$-$5.2\%\!\sqrt{1 \tn{MeV}/{\varepsilon}}$ (1.275~MeV), compatible with $1/\sqrt{\varepsilon}$-scaling.
These numbers are within a factor $\times 2$ of the $2.5\%\!\sqrt{1 \tn{MeV}/{\varepsilon}}$ obtained for X-rays in section \ref{Eres}, that deviates from the intrinsic energy resolution of the Micromegas readout at 10 bar: $1.4\%\!\sqrt{1 \tn{MeV}/{\varepsilon}}$.
A plausible explanation for the deterioration of the calorimetric response to X-rays
has been attributed to the low S/N and non-uniformity of the readout. However, the energy resolution
obtained for MeV-electron tracks suffers from yet another $\times 2$, for which it seems that a different source is needed.

 \begin{figure}[ht!!!]
\centering
\includegraphics*[width=9.0cm]{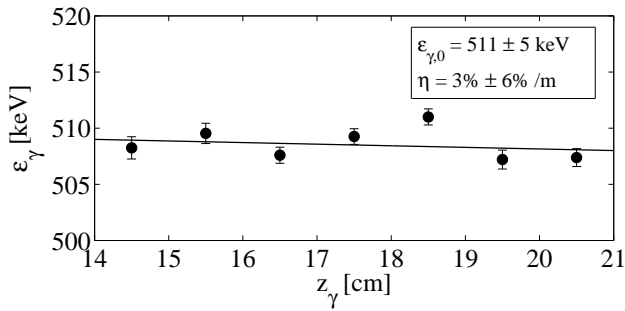}
\caption{Position of the 0.511-MeV peak as a function of the drift distance for 80 V/cm/bar and 10 bar.}
\label{Attach2}
\end{figure}

It must be noted that the positions of the 0.511 MeV and 30 keV energy peaks are well reproduced after X-ray calibration within 2\% (Figs. \ref{Attach}, \ref{Attach2}), showing little evidence for attachment or strong additional recombination besides that observed for X-rays. A marginal correlation of the reconstructed energy with the track orientation and the number of pixels fired was observed,
that may be an indication of an imperfect calibration, however it was just too small to be easily correctable,
and clearly does not account for the observed deterioration. Very modest, $<0.5\%$, improvements could be achieved by selecting specific regions along or across the chamber. The main explanation in hand comes from a relatively large system cross-talk originated during signal transmission ($1/20$-$1/30$), that is characterized in the present case by perfectly unipolar signals, and that appears synchronized with the event's largest pulses.\footnote{The best explanation for this anomalous cross-talk is the presence of $\lesssim 10k\Omega$ resistive paths or dielectric shunting conductances between some of the traces (either at the cables or at the footprints), something that had been qualitatively identified earlier. The complexity associated to the unambiguous identification of a cross-talk event, due to the presence of net charge, discouraged us from the production and study of detailed cross-talk maps.} A soft anti-coincidence condition in a $\pm 4\mu$s ($\pm 5$mm) window around the time of the event allows to suppress them when they appear isolated, however
it is virtually impossible to avoid their contribution if they are produced onto the pixels
already having a signal above or just near threshold. A ball-park estimate treating cross-talk as correlated noise with a flat probability
in the range 0-1/20 would yield 3.5\% additional peak broadening, while an average of 1.7 pixels showing cross-talk would plausibly explain the observed discrepancy between X-rays and extended electron tracks.

\subsection{Track topology}

\begin{figure}[hb!!!]
\centering
\includegraphics*[width=6.2cm]{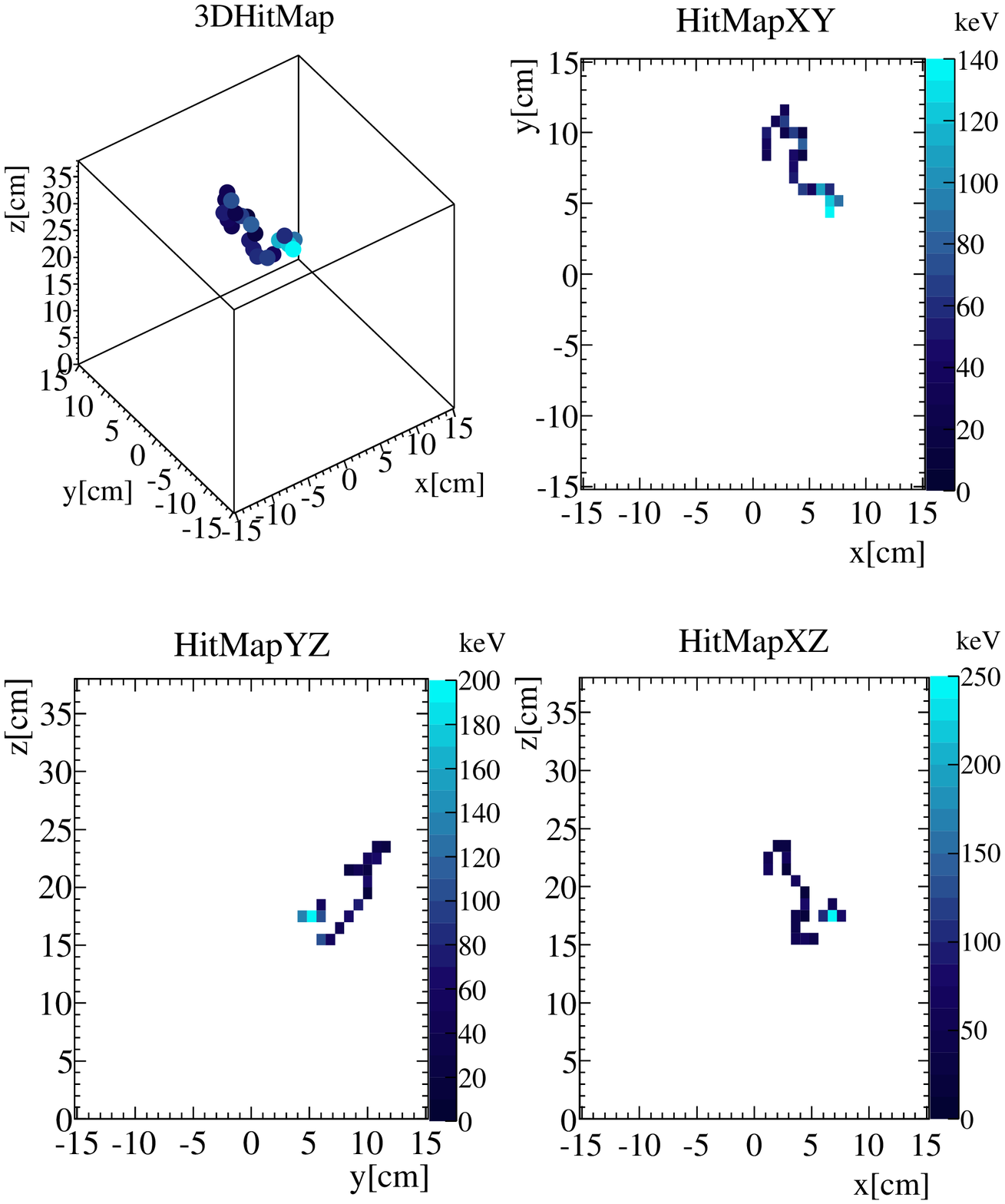}
\includegraphics*[width=6.2cm]{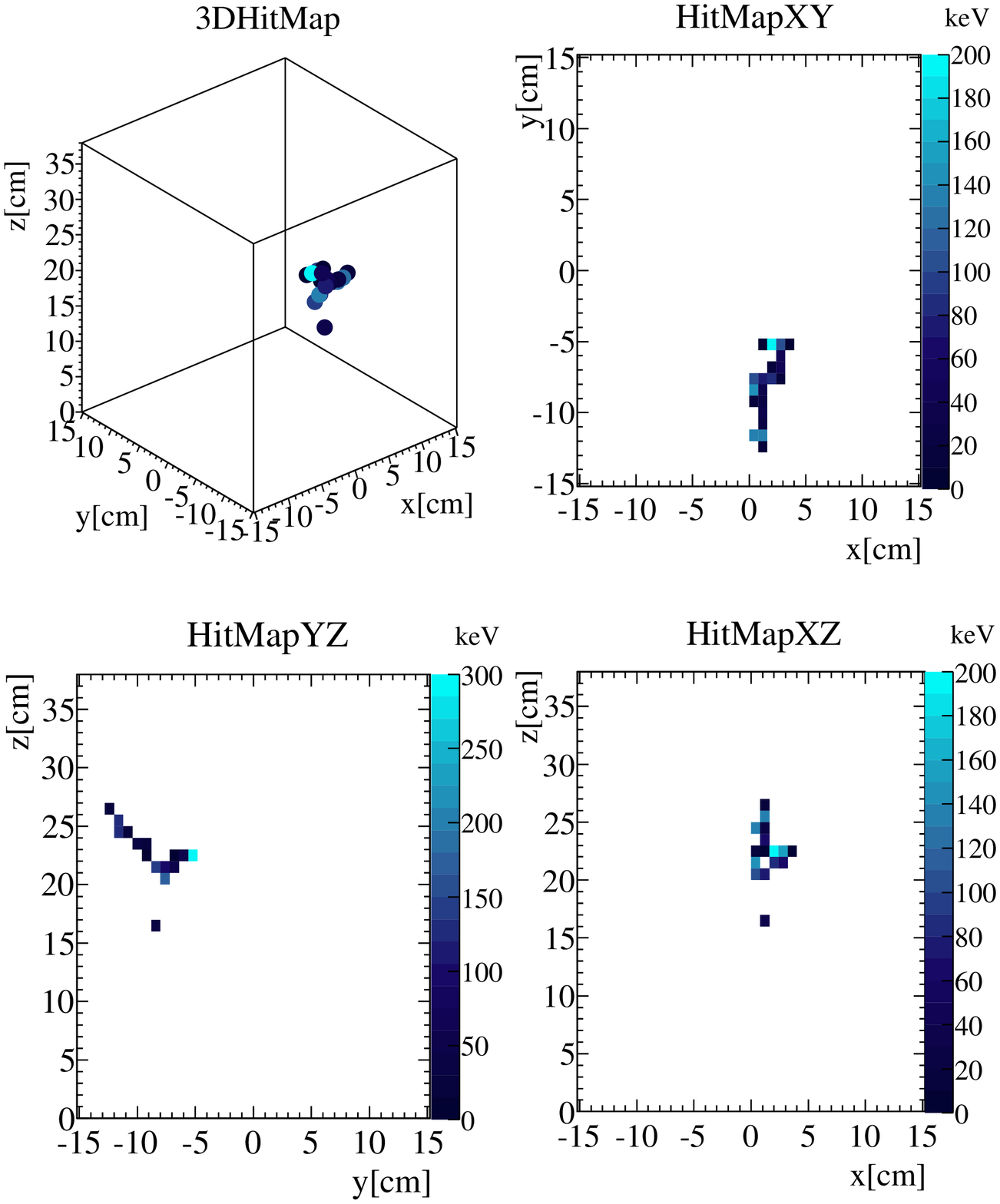}
\includegraphics*[width=6.2cm]{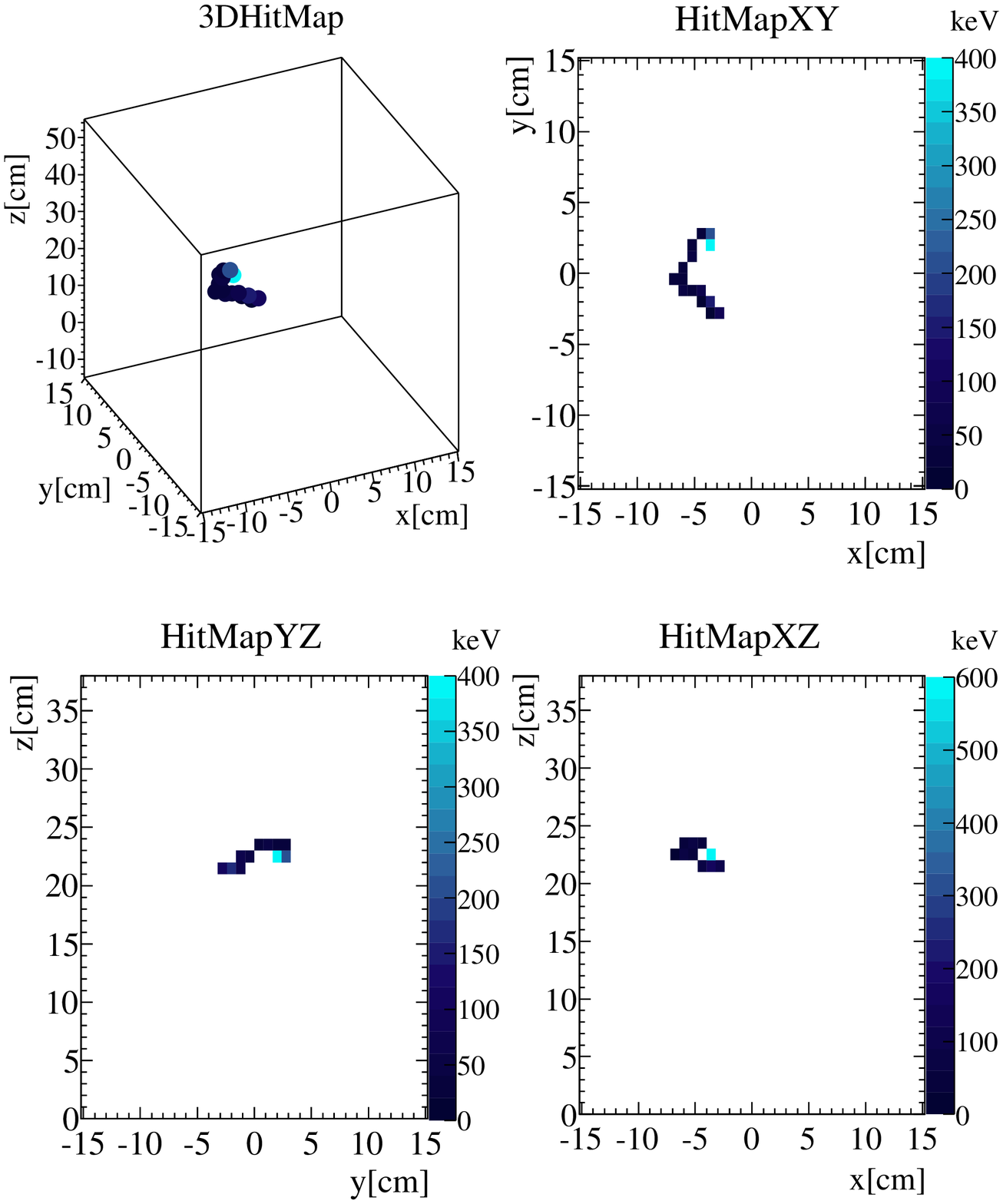}
\includegraphics*[width=6.2cm]{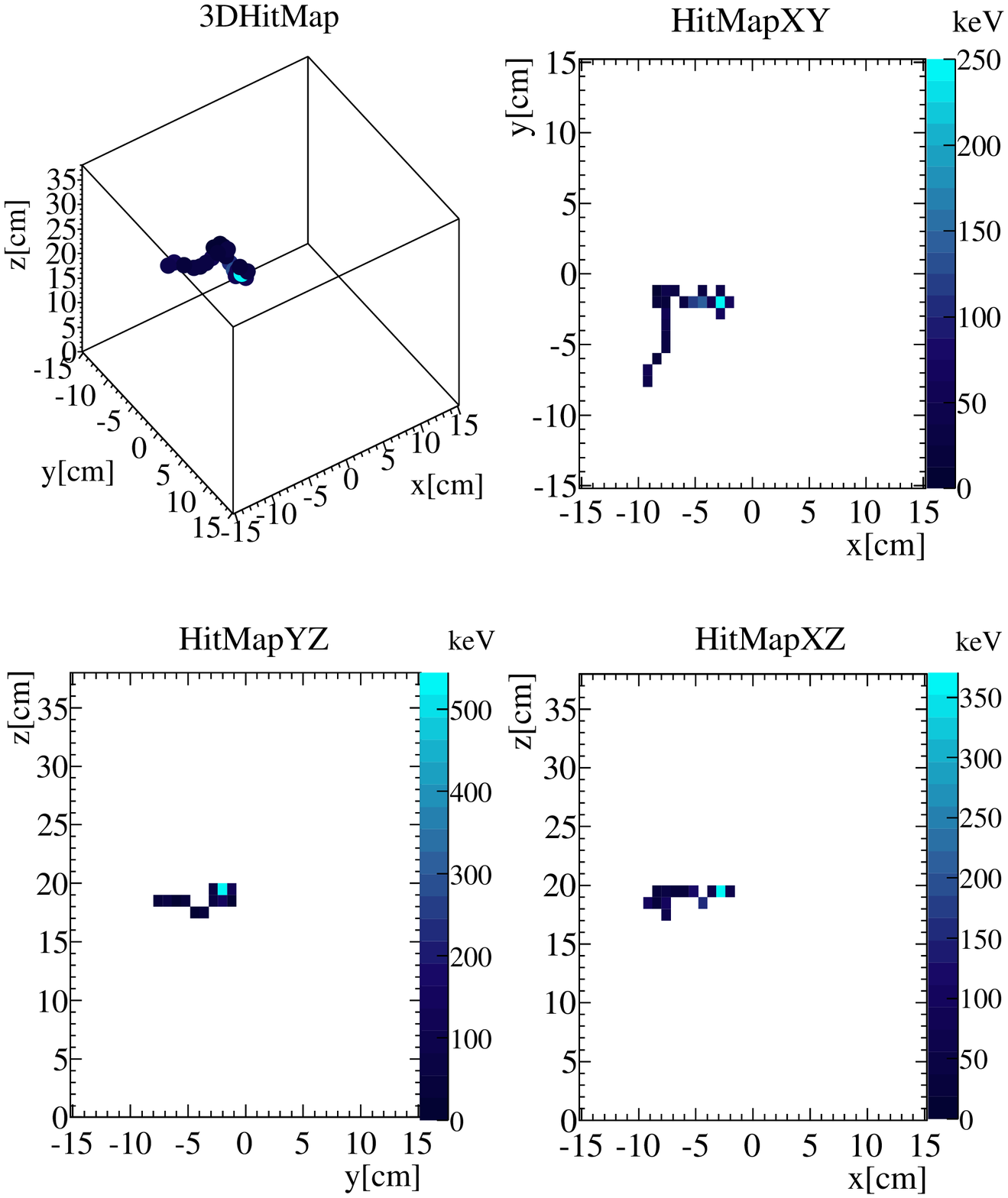}
\includegraphics*[width=6.2cm]{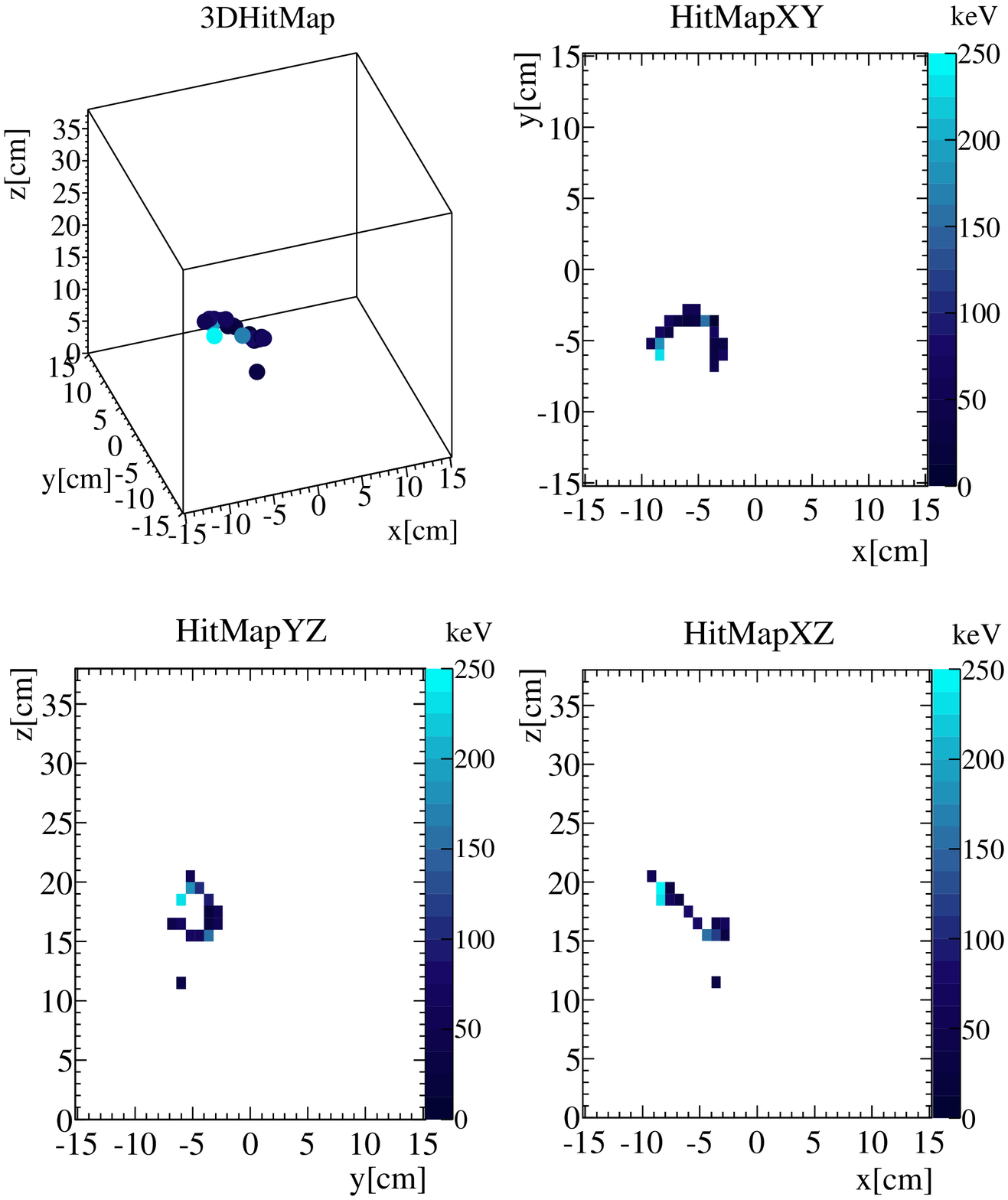}
\includegraphics*[width=6.2cm]{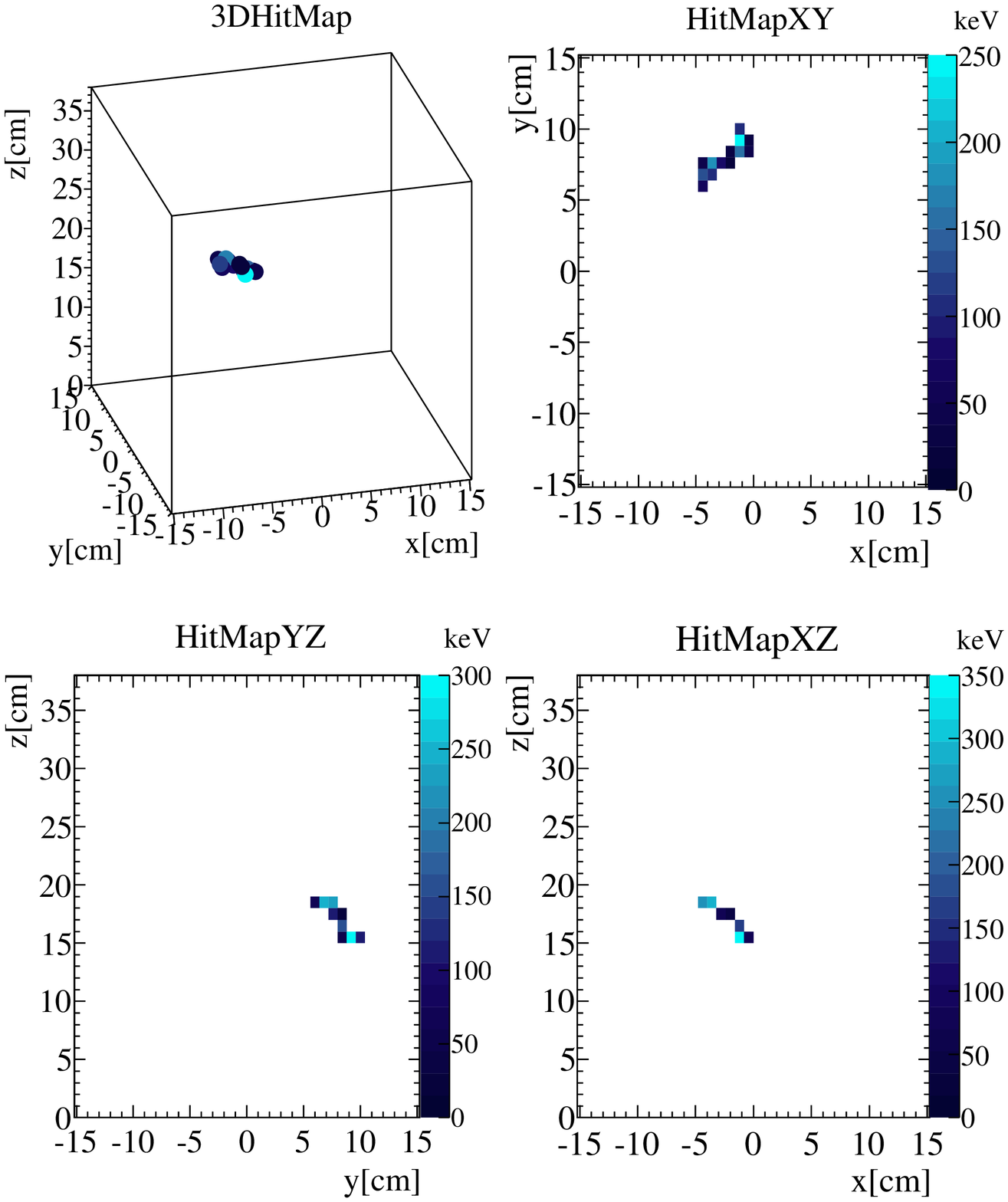}
\caption{3D reconstruction and 2D projections of several hand-picked electron tracks in a $\pm 50$ keV region around the 1.275 MeV peak, shown with a voxelization of 8 mm $\times$ 8 mm $\times$ 10 mm ($x$, $y$, $z$). Upper 2 figures for each event share color bar.}
\label{1200}
\end{figure}

Besides the calorimetric function, an essential feature for $\beta\beta0\nu$ applications is the accurate reconstruction of the electron track end-point (blob). This is illustrated in Fig. \ref{1200} through several hand-picked single-track events in the energy region around 1.275 MeV.
In this (biased-by-selection) sub-sample, that roughly describes the behaviour of
about 1/3 of the events in this energy region, the end-blob is visible to the naked eye. The remaining 2/3
consist of cosmic rays, highly bent tracks and multi-Compton events. The selected energy approaches the most probable electron
 energy in the neutrino-less decay mode ($Q_{\beta\beta}/2$), hence effectively indicating how `half' of the expected
 signal from the long sought decay looks like.

\begin{figure}[ht!!!]
\centering
\includegraphics*[width=7.0cm]{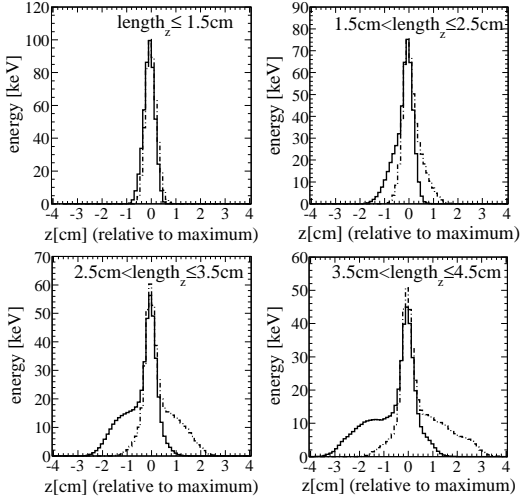}
\caption{Average energy loss profile for $0.511$ MeV-electron tracks of different lengths along the $z$ direction (1.2 mm bin). Continuous and dashed lines represent upgoing and downgoing electrons, respectively.}
\label{DeltaZs}
\end{figure}

The main practical limitation for recognizing a double-blob $\beta\beta0\nu$ event is the inability to adequately `un-fold' the tortuous electron tracks into a straight line where the blob(s) cannot be mistaken. If no attempt is made to quantify the final figure of background/signal suppression, the problem can be analyzed through a sub-sample of tracks selected to be straight along a particular spatial direction.
As will be immediately shown, 0.511 MeV tracks are long enough to allow a separation of the mip and blob region,
and so their analysis universally describes the final part of any electron of arbitrary energy in the present experimental conditions.
Provided the system's native voxelization is 8mm$\times$8mm$\times$1.2mm we focus
on the $z$ coordinate due to its higher resolution: Fig. \ref{DeltaZs} shows the superposition of all event pulses,
aligned with respect to the position of its maximum and divided by the number of events, for different track lengths along the $z$-coordinate.

As expected, the Bragg peak can be distinctly identified in extended tracks, with a blob to mip energy ratio of 5. The blob width is 2.5 mm-$\sigma$, being only partly explained by the electronics response function (1.4 mm), and with negligible contributions from binning (0.3 mm) and diffusion (0.55 mm). This sets the physical scale of the electron blob at about 2~mm-$\sigma$ (for 10 bar Xenon) implying that in present conditions not just the blob but its details start to be resolvable.

\begin{figure}[h]
\centering
\includegraphics*[width=8cm]{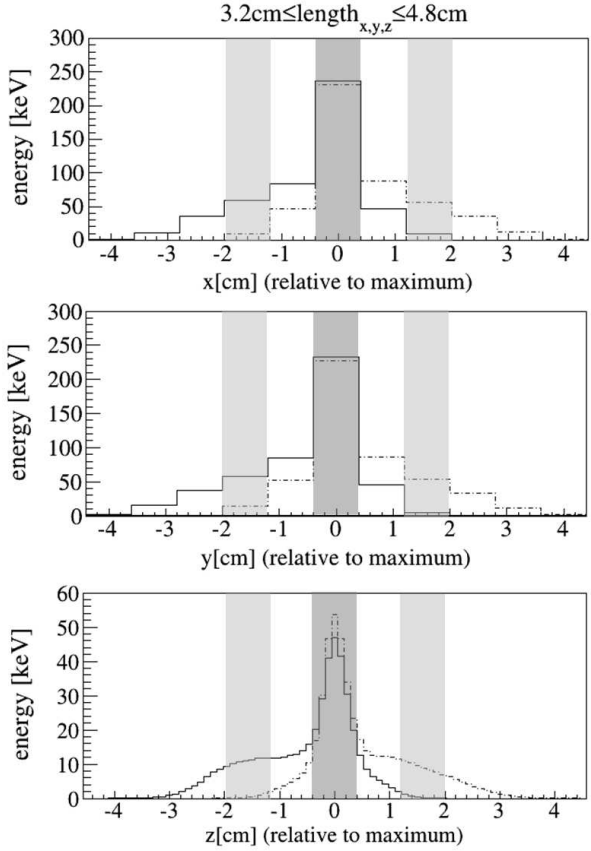}
\caption{Average energy loss profile for 0.511 MeV tracks selected to be long (3.2-4.8cm) along any of the 3 spatial projections. Binning is 8 mm for $x$, $y$ projections and 1.2 mm for $z$, following the native TPC voxelization. Continuous/dashed lines represent (from top to bottom) right/left, forward/backward, up/down -going tracks. Bands implement the definition of blob and mip region used in text.}
\label{DeltaXYZ}
\end{figure}

\begin{figure}[h]
\centering
\includegraphics*[width=8cm]{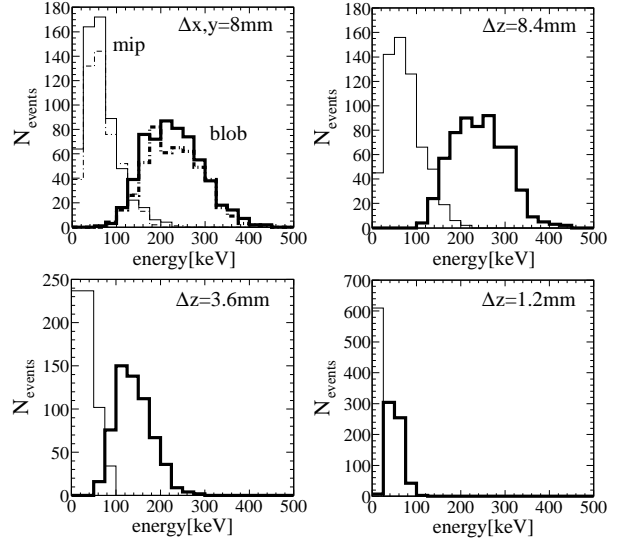}
\caption{Event-by-event blob and mip energy distributions obtained from a selection of `long' tracks (i.e. 3.2cm$\leq\tn{length}_{x,y,z}\leq4.8$cm), by selecting different $\Delta x,y,z$ bands in each projection. Up: $\Delta x,y=8$mm (left) and $\Delta z=8.4$mm (right). In the left figure the spectra corresponding to tracks extended in $x$ are indicated by continuous lines, whereas they are dashed for $y$. Bottom-left: $\Delta z=3.6$mm. Bottom-right: native $z$-binning.}
\label{Blob2mip}
\end{figure}

The influence of the pixelization in track reconstruction can be easily appraised by performing the same analysis for the longest tracks in all 3 directions of space (Fig. \ref{DeltaXYZ}). Clearly, the wider pixelization (8 mm) results in a loss of information, but how important is it?: Fig. \ref{DeltaXYZ} describes the average behaviour and extension of the final energy blob in MeV-electron tracks but does not give an idea about the energy loss fluctuations. By approximately selecting 8mm-regions both around the `blob' and `mip' regions (bands in Fig. \ref{DeltaXYZ}) identification can be attempted track-by-track, resulting in the distributions shown in Fig. \ref{Blob2mip}-up for $x$ and $y$ (up-left) and $z$ (up-right). The probability to mistakenly interpret a 8~mm mip region as a blob is below 1/20 at a signal efficiency of 96\%, if a sharp cut at 150 keV is used. The situation improves for smaller $\Delta z$ intervals until the mip and blob distributions completely decouple if the native $z$ pixelization (1.2~mm) is chosen. This again suggests that the blob region is being resolved in present experimental conditions and that mm-scale tracking capabilities are an important asset. 

\onecolumn

\twocolumn
\section{Discussion}\label{disc}
\subsection{System aspects}
From the calorimetric point of view the performance of NEXT-MM 1kg/10bar Xenon TPC compares favourably with earlier results for similar systems (Table \ref{res}) with exception of the NEXT baseline technology, based on electroluminescence. The energy resolution for a charge-readout system may be decomposed according to the contributions that have been identified in this work as:

\bear
\!\!& \mathfrak{R}\! & =\! 2.35\sqrt{\sigma_{int}^2\! + \! \sigma_{p-p}^2\! + \! \sigma_{S/N}^2\! + \! \sigma_{x-talk}^2\! + \!\sigma_{cal}^2} \label{resAll}\\
\!\!& \sigma_{int} \!& = \sqrt{F + f + \mathcal{R}} ~ \sqrt{\frac{W_I}{(1-\mathcal{R})\varepsilon}}\label{intrins}
\eear
From left to right the terms in eq. \ref{resAll} correspond to the intrinsic contribution from the Micromegas readout (including effects in drift and amplification regions), gain spread in the amplification holes (experimentally manifested as pixel-to-pixel variations of the resolution after gain equalization), limited S/N and sampling frequency, cross-talk, and calibration effects. Eq. \ref{intrins} shows the canonical formula for the intrinsic energy resolution under the assumption of decoupled fluctuations, with $f$ corresponding to the variance to gain-squared ratio of the multiplication process. Once $f$ is obtained from microscopic simulation, this formula describes small Micromegas X-ray data to good accuracy \cite{RefElisa}. Additional physical limitations to eq. \ref{intrins} for MeV-electron tracks can be reasonably discarded: i) Recombination for 0.511 MeV tracks is from present data compatible with that observed for X-rays $\mathcal{R}_{511\gamma}/\mathcal{R}_x\sim 0.98 \pm 0.02$ while a small effect is present at 1.275 MeV $\mathcal{R}_{1275\gamma}/\mathcal{R}_{511\gamma} = 0.97\pm 0.01$; ii) physical characteristics of the track like bremsstrahlung \cite{DaveTPC} are determined by the amount of Xenon and therefore any contribution to the energy resolution is largely constrained by the most accurate measurements in table \ref{res}; iii) effects related to the track orientation and charge density in the amplification holes were too small to be isolated in this analysis.

From the previous arguments it may be assumed that the measured energy resolution is dominated by instrumental limitations and not by the different physical behaviour of X-ray and $\gamma$-tracks. Several sources of improvement can be identified: i) according to \cite{AFTER} noise minimization down to ENC$=2000$e$^-$ is feasible with a reduced cable capacitance down to 50 pF (C$_{pix}$+C$_{cable}\simeq180$ pF in this system \cite{comm}) while increased sampling frequency ($\times 4$) and cross-talk optimization in cables and connections should suppress the most important contributions to eq. \ref{resAll}: $\sigma_{S/N}$ and $\sigma_{x-talk}$. The implementation of well targeted QC procedures in connection with hole-pattern optimization, geometrical gain compensation \cite{GioComp} or increased fabrication accuracy could bring further improvements in the magnitude and uniformity of the operating gain. This will help at further reducing the terms $\sigma_{S/N}$ and $\sigma_{p-p}$, but it remains to be demonstrated.

\begin{table}[h]
  \centering
  \begin{tabular}{|c|c|c|c|c|}
     \hline
      & $\mathfrak{Res}$(1 MeV) & energy & $P$ & gas\\
     \hline
     Gotthard\cite{Gotthard}   & 10.6\%           & 511 keV          & 5  bar  & Xe/CH$_4$\\
     EXO200\cite{EXO200}       & 8.5\%           & 1000 keV(int.)& liquid  & Xe   \\
     NEXT\cite{Lorca}          & 1.15\%(0.97\%)   & 511(30) keV      & 10 bar  & Xe\\
     this work                 & 4.5-5.5\%(2.5\%) & 511(30) keV      & 10 bar  & Xe/TMA \\
     this work                 & 4.4$\pm0.9$\%-5.2$\pm1.3$\% & 1275 keV      & 10 bar  & Xe/TMA \\
     \hline
     small MM\cite{XeTMA}      & 1.4\%           & 22 keV           & 10 bar  & Xe/TMA \\
     small MM\cite{PureXe}     & 4.8\%            & 22 keV           & 10 bar  & Xe     \\
     Fano limit \cite{DaveTPC} & 0.45\%           & -                & gas     & Xe     \\
     \hline
   \end{tabular}
  \caption{Energy resolution ($1/\sqrt{\varepsilon}$-extrapolated to 1 MeV for comparison) in Xenon-based detectors and for different electron energies. For reference, values measured in state of the art microbulk Micromegas under localized irradiation and the Fano factor limit are shown. They approximately represent intrinsic limits to charge-amplification readouts and electroluminescence, respectively.}\label{res}
\end{table}

Besides the good energy resolution observed for 0.511 MeV electrons these type of readouts have remarkable strengths, like the easiness and scalability of the readout segmentation, simplicity and high fidelity to primary ionization. The chosen voxelization of 8 mm$ \times 8$ mm$ \times 1.2$ mm provides a clean separation between blob and mip regions, largely unaffected by diffusion even at the 1m-scale. The detector can work stable for 100 live-days without signs of deterioration except for a mild damage rate extrapolable to about 1\%/year or less due to conditioning effects.

\subsection{Penning-Fluorescent behaviour} \label{PF}

\begin{figure}[h!!!]
 \centering
 \includegraphics*[width=8cm]{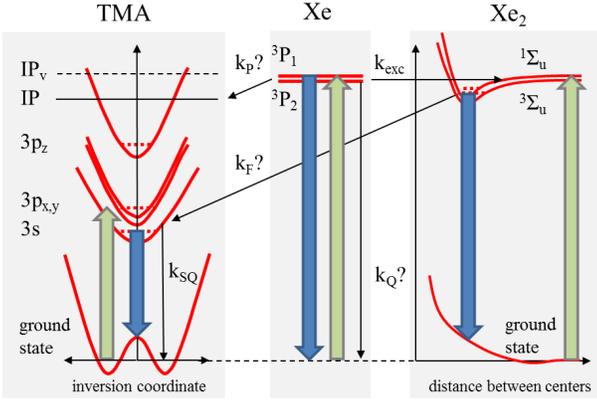}
 \caption{Main processes involved in a `Penning-Fluorescent' Xenon-TMA mixture. Approximate potential surfaces are represented as a function of the principal coordinate (for details the reader is referred to \cite{Cardoza,Lee,XeLast}). Energy levels display the correct scale: for TMA $E_{3s}= 5.460$ eV, $E_{3p_{x,y}} = 6.161$ eV, $E_{3p_{z}} = 6.229$ eV, adiabatic ionization potential $IP=7.74$ eV, vertical ionization potencial $IP_{_\tn{v}}=8.44$ eV, $\lambda_{scint}\sim 300$ nm; for Xe $E_{^3\!P_2}=8.31$ eV, $E_{^3\!P_1}=8.43$ eV; for Xe$_2$ $E_{^3\!\Sigma_u^+}=7.92$ eV, $E_{^1\!\Sigma_u^+}=8.05$ eV, $\lambda_{scint}\sim 172$ nm. Only the most obvious path-ways are shown (arrows): Penning and fluorescent transfer rates (characterized by $k_{P}$ and $k_{F}$) as well as Xe$^*$ quenching rates in the presence of TMA ($k_{Q}$) remain unknown at the moment. An estimate of the Penning probability ($r_{P}$) as a function of $k_{P}$, $k_{Q}$ and the rate of excimer formation ($k_{exc}$) is given later in text.}
 \label{TMAXeLevels}
 \end{figure}

Indirect evidence for Penning-transfer reactions in Xe-TMA mixtures has been obtained in wire-chambers \cite{Ramsey}, GEMs \cite{Fraga}, and Micromegas \cite{XeTMA}. Their scintillating properties on the other hand have been reported recently \cite{Yasu}. If considering only energy transfers between the lowest lying states,
the behaviour of a (VUV-quenched) Penning-Fluorescent mixture can be illustrated through the simplified Xe-TMA diagram in Fig. \ref{TMAXeLevels}.
The relevant reaction rates are $k_Q$ (Xe$^*$ quenching by TMA), $k_P$ (Penning transfer rate), $k_F$ (fluorescent transfer rate) and  $k_{SQ}$ (TMA self-quenching rate), of which the first three are unknown. In this discussion we interpret these rate constants as effective parameters, keeping in mind that more complex behaviours may ensue, e.g., with fluorescence sequentially following Penning.

First, it must be noted that a lower bound to the primary scintillation expected from a Xenon-TMA mixture can be derived from studies at low pressure, by considering only direct TMA excitation/scintillation and self-quenching. Hence, we conservatively omit the presence of Xe$^*$, Xe$_2^*\rightarrow$TMA$^*$(3s) transfers ($k_F\simeq0$), and the quenching of TMA excited states by Xenon species, following \cite{ObiXe}. Such a bound results in:

\beq
S_1(TMA) = \frac{1}{W_{sc}} \frac{c}{\tau_{3s}(P \cdot c \cdot k_{SQ} + 1/\tau_{3s})} \label{S1direct}
\eeq
as a function of concentration ($c$) and pressure ($P$). By considering for TMA a typical value $W_{sc}=30\pm{10}$ eV ($\simeq 5.5 \times E_{3s}$), the measured lifetime of the 3s state ($\tau_{3s}=44$ ns) and the TMA self-quenching rate ($k_{SQ}= 1.08$ ns$^{-1}$ bar$^{-1}$), evaluation of eq. \ref{S1direct} leads to S$_1$ yields up to 100's of ph/MeV, depending on pressure \cite{DiegoLessons}. Additional scintillation hangs on the ability of TMA to quickly dissipate, collisionally, a fairly large amount of internal excess energy ($\sim 2$ eV) following the Xe$^*$, Xe$_2^*\rightarrow$TMA$^*$ transfers; relaxation to the 3s state may in this way happen while avoiding the dissociative pathways \cite{ObiXe}. Although with limited efficiency, this transfer process is strongly hinted by the measurements in \cite{Yasu}.

To the extent that energy transfers occur with the Xe$^*$ states, Penning reactions are energetically viable:\footnote{Other channels are generally present \cite{Ozkan}, but they do not change the general argument followed here.}

\beq
\tn{Xe}^* + \tn{TMA} \rightarrow \tn{Xe} + \tn{TMA}^+ + \tn{e}^- \label{PenningEq}
\eeq
The increased charge collection associated to the presence of Penning transfers can be a beneficial agent, reducing the Fano factor \cite{Knoll} and increasing the ionization response of the admixture; in the case of charge multiplication the avalanche fluctuations are reduced at any given gain \cite{RobSta}.
By comparing Magboltz simulations with data obtained from charge amplification in Micromegas, an approximate functional behaviour of the Penning probability $r_P$ has been derived in \cite{RefElisa}, as:

\beq
r_P = \frac{c P k_P}{(1-c)^2P^2k_{exc} + c P (k_{P} + k_{Q}) + 1/\tau^*}
\eeq
where $\tau^*$ represents an effective life-time of the state from which Penning takes place.
Maximum Penning transfer rates in the range 10-30\% are hence obtained, decreasing with pressure.

\subsection{Outlook}

It has been shown that the use of trimethylamine as VUV-quencher in Xenon results in a 20-fold reduction of the transverse diffusion relative to pure Xenon. Besides allowing the stable operation of microbulk Micromegas at 10 bar, Xe-TMA exhibits substantial primary scintillation above $250~$nm, at the level of $\sim\!100$'s ph/MeV \cite{Yasu}. The energy resolution obtained for 0.511 MeV $\gamma$'s is at least $\times 2$ better than the one obtained earlier with wire chambers, with Xe-TMA reducing electron diffusion in factors of $\simeq 2$ in all 3 space dimensions relative to, e.g., Xe-CH$_4$.

From the above considerations it may seem that, if present radiopurity levels of microbulk Micromegas (0.4~Bq/m$^2$) can be reduced to the levels expected from the raw materials alone, re-sizing the present concept `as is' could greatly improve on earlier limits obtained for $\beta\beta0\nu$-decay in high pressure Xenon in \cite{Gotthard}. In this scenario, the optimum pressure is undoubtedly an essential variable to be considered. Gains of 2000 were shown for small sensors at 5 bar in \cite{XeTMA} as compared to 400 at 10 bar. A maximum gain of 200 could be achieved for the present $700$cm$^2$-readout system, limiting the energy resolution.

Despite the positive comparison with earlier attempts in charge-mode readouts, it remains unclear to this collaboration how such a technological approach could outperform an electro-luminescence option, specially if more conventional low-diffusion mixtures at tolerable VUV-quenching rates do exist, as it seems \cite{DiegoTPC}. On the other hand, although much more challenging, the studies performed on Xe-TMA show that the existence of a magic `Penning-Fluorescent' mixture with high Penning and (wavelength-shifted) fluorescence yields cannot be discarded out of hand. Work in this direction will continue.

\section{Conclusions}\label{conc}

A Time Projection Chamber (NEXT-MM) housing 1.1 kg of Xenon in its fiducial volume and instrumented with a microbulk Micromegas readout plane (manufactured `a la' GEM) has been operated for 100 live-days in a 10 bar Xenon/trimethylamine `Penning-Fluorescent' gas mixture (99/1). The TPC ran continuously and without the need of dedicated shifts, showing a Micromegas damage rate at the halting of operation extrapolable to 1\%/year. A seemingly reducible 5\% damage was observed during system commissioning. The loss of exposure due to the instabilities introduced by the observed rate of (non-damaging) micro-discharges in the readout was quantified to be less than 0.05\%.

We have detailed the calibration procedures and fully characterized the TPC response with Xenon X-rays. The energy resolution (FWHM) ultimately achieved on the full fiducial volume was 14.6\% at 30 keV, with
contributions from the limited S/N, sampling frequency and non-uniformity of the readout plane explaining the deterioration with respect to results obtained earlier in small setups (9\%). At the $Q_{\beta\beta}$ of $^{136}$Xe, and assuming $\sqrt{\varepsilon}$ scaling, the calorimetric response to 0.511~MeV and 1.275~MeV electron tracks anticipates FWHM energy resolutions from 3.2\%(for the best sector) to 3.9\%(full TPC). We have proposed concrete ways to further mitigate the damage rates and to improve the readout. They could bring the energy resolution to the level obtained for X-rays ($1.6\%$ at $Q_{\beta\beta}$), or even to the intrinsic value obtained for small sensors ($1.0\%$ at $Q_{\beta\beta}$).

A simplified topological analysis carried out with the system's native voxelization of 8mm$\times $8mm$\times$1.2mm (${x}\times{y}\times{z}$) indicates a strong enhancement of the mip-to-blob identification capabilities in the $z$-direction to the extent of completely decoupling the mip and blob energy distributions for a sub-sample of 600 selected tracks. Due to the very low electron diffusion measured for the mixture (at the scale of 1~mm for 1~m drift), and the easiness at increasing the readout granularity, the technology introduced thus offers the possibility of mm-accurate true-3D reconstruction of MeV-electron tracks on large detection volumes and at high pressure. In particular, it was shown that the electron end-point or `blob' (2mm-$\sigma$ at 10 bar) could be well resolved with little ambiguity.

Taking the performance of the NEXT-MM TPC at face value, the scaling of the present system `as is' could supersede earlier attempts to find the $\beta\beta0\nu$ decay with wire-chambers. In order to be competitive with a technology based on electroluminescence, a solution to the `$T_o$-problem', as well as an enhanced calorimetric response, seem to be needed. Also, by virtue of their low material budget and radiopure materials, Microbulk micromegas hold a promise for ultra-high radiopurity, therefore efforts to improve the present bounds of around 0.4 Bq/m$^2$ for the $^{238}$U and $^{232}$Th chains should continue.

\ack
The author acknowledges Steve Biagi and Francisco
Fraga for many useful discussions on the theoretical and experimental aspects of TMA, as well as `Mar de Cadiz'
for strong support during the final stage of the manuscript. The role of CERN's workshop
under R. de Oliveira was instrumental at achieving the high quality necessary for operation
of single-stage micro-hole gaseous amplification devices at the 1000 cm$^2$ scale. The great benefit
stemming from the discussions and encouragement of our
RD51 colleagues can be hardly understated.

The NEXT collaboration acknowledges funding support from the following agencies and institutions:
European Research Council under Advanced Grant 339787-NEXT and Starting Grant 240054-TREX,
Spanish Ministerio de Econom\' ia y Competitividad under grants Consolider-Ingenio
2010 CSD2008-0037 (CUP) and CSD2007-00042 (CPAN), contracts FPA2008-03456 and
FPA2009-13697; Portuguese Funda\c{c}$\tilde{\tn{a}}$o para a Ci\^{e}ncia e a Tecnologia;
European FEDER under grant PTDC/FIS/103860/2008; US Department Of Energy under contract DE-AC02-05CH11231.


\begin{thebibliography}{00}
\bibitem{AzeCC} C.D.R. Azevedo et al.,  Nucl. Instr. Meth. A, 732(2013)551.
\bibitem{Matsu} Y. Matsuoka et al., JINST 10(2015)C01053PET.
\bibitem{HARPO} D. Bernard, Nucl. Instr. Meth. A, 729(2013)765.
\bibitem{Black} J. K. Black, J. Phys. Conf. Ser, 65(2007)012005.
\bibitem{NEXT}  \'Alvarez~V. et al. (NEXT Collaboration), JINST, 7(2012)T06001
\bibitem{AGATA} S. Akkoyun, Nucl. Instr. Meth. A, 668(2012)26.
\bibitem{Alzakhov} G. D. Alkhazov et al., Nucl. Instr. Meth. 48,1(1967)1.
\bibitem{Gotthard} R. Luescher et al., Phys. Lett. B, 434(1998)407.
\bibitem{DaveTPC} D. Nygren, Nucl. Instr. Meth. A 603(2009)337.
\bibitem{LOI} F. Gra\~nena et al, NEXT letter of intent to Laboratorio Subterraneo de Canfranc, arXiv:0907.4054.
\bibitem{NEXT-DBDM} \'Alvarez~V. et al. (NEXT Collaboration), Nucl. Inst. and Meth. A, 114(2013)101.
\bibitem{Policarpo} A. J. P. L. Policarpo, Physica Scripta, 23(1981)539.
\bibitem{NEXT_TDR} \'Alvarez~V. et al. (NEXT Collaboration), JINST 7(2012)T06001.
\bibitem{DEMO} \'Alvarez~V. et al. (NEXT Collaboration), JINST, 8(2013)P04002.
\bibitem{Lorca} D. Lorca et al., (NEXT Collaboration), JINST, 9(2014)P10007.
\bibitem{JJBaTa} J. J. Gomez-Cadenas, \textit{The NEXT experiment} arXiv:1411.2433, submitted to Nucl. Phys. B(Proc. Suppl.).
\bibitem{pdg} K.A. Olive et al. (Particle Data Group), Chin. Phys. C, 38, 090001 (2014).
\bibitem{DavePF} D. Nygren, J. Phys. Conf. Ser., 309(2011)012006.
\bibitem{HerreraReco} D. C. Herrera, PoS(TIPP2014)057.
\bibitem{Sinclair} D. Sinclair et al., J. Phys. Conf. Ser. 309(2011)012005.
\bibitem{DiegoTMA} V. Alvarez et al. (NEXT Collaboration), JINST 9(2014)C04015.
\bibitem{XeTMA} S. Cebrian et al., JINST 8 (2013) P01012.
\bibitem{Ramsey} B. D. Ramsey, P. C. Agrawal, Nucl. Instr. Meth., A 278(1989)576.
\bibitem{Grosjean} D. Grosjean, P. Bletzinger, IEEE J. Quantum Elect. QE-13, 11(1977)898.
\bibitem{Tannembaum} Eileen Tannembaum, Esther M. Coffin, Anna J. Harrison, J. Chem. Phys. 21(1953)311.
\bibitem{Halpern} Arthur M. Halpern, Mary Jo Ondrechen, Lawrence D. Ziegler, J. Am. Chem. Soc. 108(1986)3907.
\bibitem{Obi1} Yutaka Matsumi, Kinichi Obi, Chem. Phys. 49(1980)87.
\bibitem{Cureton}  Clifford G. Cureton et al., Chem. Phys. 63(1981)31.
\bibitem{ObiXe} Kinichi Obi and Yutaka Matsumi, Chem. Phys. 49(1980)95.
\bibitem{Cardoza} Job D. Cardoza et al., J. Phys. Chem. A, 112(2008)10736.
\bibitem{Koehler} H. A. Koehler et al. Phys. Rev. A, 9,2(1974)768.
\bibitem{muBulk} S. Andriamonje et al., JINST 5(2010)P02001.
\bibitem{HecRP} T. Dafni et al., Astropart. Phys. 34(2011)354.
\bibitem{Topo} J. Mart\'in-Albo et al. (NEXT Collaboration), \textit{ Sensitivity of NEXT-100 to neutrinoless double beta decay }, to be submitted to JHEP.
\bibitem{comm} V. Alvarez et al. (NEXT Collaboration), JINST 9(2014)P01301.
\bibitem{AFTER} Baron~P. et al., IEEE Trans. Nucl. Sci., 55(2008)1744.
\bibitem{DiegoRev} Diego Gonzalez-Diaz, Archana Scharma, JINST 8(2013)T02001.
\bibitem{CASTres} S. Aune et al., JINST 8 (2013) C12042.
\bibitem{Magboltz} S. F. Biagi., http://magboltz.web.cern.ch/magboltz/.
\bibitem{Yasu} Y. Nakajima, \textit{Measurement of scintillation and ionization yield with high-pressure gaseous mixtures of Xe and TMA for improved 0nu double beta decay and dark matter searches}, arXiv:1505.03585 [physics.ins-det].
\bibitem{RefElisa} E. Ruiz-Choliz, D.Gonzalez-Diaz, A. Diago et al., \textit{Modelling the behaviour of microbulk Micromegas in Xenon-trimethylamine gas mixtures}, arXiv:1506.05077[physics.ins-det], submitted to Nucl. Instr. Meth. A.
\bibitem{GioComp} Y. Giomataris, Nucl. Instr. Meth. , 419(1998)239.
\bibitem{EXO200} M. Auger et al., Phys. Rev. Lett. 109(2012)032505.
\bibitem{Fraga} F. A. F. Fraga et al., Nucl. Instr. Meth. A, 513(2003)379.
\bibitem{DiegoLessons} Diego Gonzalez-Diaz et al., \textit{Lessons from the `Penning-Fluorescent' TPC and prospects}, arXiv:1504.03640 [physics.ins-det].
\bibitem{Lee}    J. B. Nee et al., Chem. Phys. Lett., 318(2008)402.
\bibitem{XeLast} Di Zhu, Xion Zhang, Hiroshi Kajiyama, J. Appl. Phys. 112(2012)033304.
\bibitem{Ozkan} O. Sahin et al., JINST 5(2010)P05002.
\bibitem{Knoll} G. F. Knoll, \textit{Radiation detection and measurement}, John Wiley\&Sons, 2010.
\bibitem{RobSta} H. Schindler, S. F. Biagi, R. Veenhof, Nucl. Instr. Meth., A 624(2010)78.
\bibitem{DaveReco} D. R. Nygren, J. Phys. Conf. Ser. 460(2013)012006.
\bibitem{PureXe} C. Balan et al. JINST 6(2011)P02006.
\bibitem{DiegoTPC} D. Gonzalez-Diaz (NEXT collaboration), \textit{NEXT-generation HP Xe-TPCs for the NEXT-$\beta\beta0\nu$ experiment}, presented at the 7th symposium for low-energy rare event detection.
\end{thebibliography}
\end{document}